\documentclass[prb,twocolumn]{revtex4-1}
\usepackage{amsmath, amsthm, amsfonts, amssymb, graphicx}

\newcommand{\half}{\tfrac{1}{2}}
\newcommand{\ev}[1]{\langle #1\rangle}
\newcommand{\ket}[1]{| #1 \rangle}
\newcommand{\bra}[1]{\langle #1 |}
\newcommand{\Tr}{\mathrm{Tr}\,}
\newcommand{\Id}{\mathrm{1\!\!1}}
\newcommand{\Itilde}{{\tilde{I}}}
\newcommand{\Ibar}{{\overline{I}}}
\newcommand{\Inn}{{I^{[j]}}}
\newcommand{\Ipnn}{{{I'}^{[j']}}}
\newcommand{\Innp}{{{I}^{[j']}}} 
\newcommand{\xhat}{\hat{x}}
\newcommand{\yhat}{\hat{y}}

\newcommand{\dn}{\downarrow}
\newcommand{\up}{\uparrow}

\begin{document}

\title{A real-space renormalization-group calculation for the \\
quantum $\mathbb{Z}_2$ gauge theory on a square lattice}

\author{Steve T. Paik}
\email[]{paik\_steve@smc.edu}
\affiliation{Santa Monica College,\\ Santa Monica, CA 90405}

\date{\today}

\begin{abstract}
We revisit Fradkin and Raby's real-space renormalization-group method to study 
the quantum $\mathbb{Z}_2$ gauge theory defined on links forming a 
two-dimensional square lattice. Following an old suggestion of theirs, 
a systematic perturbation expansion developed by Hirsch and Mazenko is used to
improve the algorithm to second order in an intercell coupling, thereby 
incorporating the effects of discarded higher energy states. A careful
derivation of gauge-invariant effective operators is presented in the
Hamiltonian formalism. Renormalization 
group equations are analyzed near the nontrivial fixed point, reaffirming old 
work by Hirsch on the dual transverse field Ising model. In addition to 
recovering Hirsch's previous findings, critical exponents 
for the scaling of the spatial correlation length and energy gap in the 
electric free (deconfined) phase are compared. 
Unfortunately, their agreement is poor. 
The leading singular behavior of the ground state energy density is examined
near the critical point: we compute both a critical exponent and estimate a
critical amplitude ratio.
\end{abstract}

\maketitle

\section{Introduction}

We study the quantum Hamiltonian for a two-dimensional $\mathbb{Z}_2$ gauge 
theory on a square lattice using a real-space renormalization-group method. The
method, due to Fradkin and Raby, is a gauge-invariance-preserving block-spin 
algorithm with length rescaling factor two. A variational approximation is made
for the ground state of the theory and the Hilbert space is thinned so that 
low-energy states and long-distance correlations are preserved. Despite the 
crudeness of the truncation, we demonstrate, without recourse to duality, that 
spatial correlations decay exponentially in the electric free, or deconfined,
phase. 

The quantum Hamiltonian may be obtained from classical statistical mechanics 
by starting with a euclidean three-dimensional $\mathbb{Z}_2$ gauge theory on a 
lattice with anisotropic couplings $\beta_t$, along a particular direction 
chosen as ``time,'' and $\beta_s$ in the orthogonal directions. When the 
partition function is expressed in terms of a transfer operator, a special 
limit exists in which the Trotter product formula allows for the transfer 
operator to be expressed as the exponential of some Hamiltonian $H$. \cite{SML}
This is an infinite-volume limit that is also highly anisotropic and requires 
$\beta_t \to \infty$ and $\beta_s \to 0$ such that $\beta_s e^{2\beta_t}$ 
remains a fixed and arbitrary dimensionless coupling. \cite{FS} 

At each link $l$ there exists spin-$\half$ operators $\vec{\sigma}_l$ obeying 
the Pauli algebra. Operators belonging to different links commute. The 
Hamiltonian is
\begin{equation}
  H = -h \sum_l \sigma^z_l - J \sum_p\Phi_p,
\end{equation}
where $\sigma^z_l$ measures the discrete electric flux running along link $l$,
and $\Phi_p = \prod_{l \in \partial p}\sigma^x_l$ measures the discrete 
magnetic flux through plaquette $p$.

Local gauge transformations are defined by operators associated to the sites
or vertices between links. At such a site $\vec{r}$ in the lattice the 
generator is
\begin{equation}
  G_{\vec{r}} = \prod_{\substack{\text{links $l$ emerging}\\ 
  \text{from $\vec{r}$}}} \sigma^z_l.
\end{equation}
$G_{\vec{r}}$ commutes with $H$.

\begin{table}
\begin{ruledtabular}
\begin{tabular}{cccc}
  order in Hirsch--Mazenko & & \\
  perturbation theory & $\nu_t$ & $\nu_s$ & $\alpha$ \\
  \hline
  first (Refs.~\onlinecite{FR,MG}) & 0.62 & 1.20 & $-1.02$ \\
  second (Ref.~\onlinecite{Hirsch} and this work) & 0.49 & 0.65 & 0.21
\end{tabular}
\end{ruledtabular}
\caption{\label{tab:critexp}
Critical exponents for the vanishing of the energy gap ($\nu_t$), spatial 
correlation length ($\nu_s$), and singularity in the ground state energy 
($\alpha$) in the Hirsch--Mazenko perturbation expansion.}
\end{table}

The renormalization-group transformation developed by Fradkin and Raby fixed an
important shortcoming of previous real-space schemes. \cite{FR}
Although blocking spin 
operators into cells makes a variational approximation to the lattice ground 
state analytically tractable, such approximations are engineered to preserve 
the low-energy spectrum without regard for spatial correlations. However, in 
the Hamiltonian formalism, time and space are treated on very different 
footings, so one generally needs to ensure that both dimensions scale equally 
under the renormalization transformation. Consequently, while the gap is 
well-approximated, (equal-time) correlations exhibit qualitatively incorrect 
behavior such as power-law decay away from criticality. Fradkin and Raby found 
that such long-range correlations may be suppressed in the ground state by 
designing the neighboring block spins to share boundary conditions. This 
prevents cells from being disconnected because the magnetic flux of one cell is 
not entirely independent of the flux of its nearest neighbor. They proved that 
correlation functions decay exponentially in the disordered phase of the
one-dimensional transverse field Ising chain. Since the same conclusion holds 
for the two-dimensional Ising model in a transverse field, Fradkin and Raby
invoke duality to argue that it must be true for the $\mathbb{Z}_2$ gauge 
theory in the electric free phase. We shall, in fact, demonstrate this directly
using the transformation relations in the gauge theory. 

Despite the qualitative success of the real-space renormalization-group
transformation, quantitative success eludes this method. In their 
analysis, Fradkin and Raby used a square block of linear size $2$ (in units of
the lattice spacing). Thus, length scales double after each iteration of the 
blocking transformation. Unfortunately, time does not scale the same way.
They find that, even at the fixed point, where one expects the rotational 
invariance of the classical three-dimensional gauge theory to be fully restored,
time increases only by a factor of $1.43$.

Improving upon the lowest order results from the block-spin programme has been
a mixed bag, leading some authors to conclude that real-space methods are, at
best, semiquantitative. \cite{Privman}
In studies of the one-dimensional transverse field Ising model
enlarging the cell size does improve the computation of critical 
exponents. \cite{JPFD} However, the convergence to exact results are slow 
and the computation ceases to be analytically tractable beyond cells containing
more than a few spins. Certainly, real-space renormalization has not been as 
successful in producing precise output in the critical regime as other 
techniques (e.g., epsilon expansion, high-temperature series, Monte Carlo, 
conformal bootstrap). The difficulty of systematically correcting the 
variational approximation also casts some doubt as to whether the procedure is 
physically reasonable---it is hard to judge the efficacy of the approximation 
based solely on the numerical proximity of exponents as this might be 
accidental.

It was suggested by Fradkin and Raby that the asymmetric scaling of space and 
time in their transformation could be remedied by applying a perturbative
formalism developed by Hirsch and Mazenko in Ref.~\onlinecite{HM}. 
In this approach virtual effects
arising from decimated degrees of freedom generate effective operators 
connecting nearby plaquettes and links. The expansion is organized around a 
parameter, $g$, such that the original results of Fradkin and Raby are obtained
at order $g^1$, and quantum fluctuations from truncated cell spectra influence 
the renormalized Hamiltonian at order $g^2$ and beyond by generating new
effective interactions. The order $g^2$ calculation has been studied by Hirsch 
in the context of the Ising model. \cite{Hirsch} In the present work we show 
how the same formalism can be applied in the $\mathbb{Z}_2$ gauge theory. We 
find the same results for critical exponents as Hirsch because our effective 
operators in the lattice gauge theory map into those of the Ising model 
according to the well-known duality transformation. The results confirm that 
the scaling of spatial correlations improves significantly in going from order
$g^1$ to $g^2$. We understand this to be due to the inclusion of more 
delocalized operators in the effective Hamiltonian. Unfortunately, the gap 
critical exponent worsens significantly. We understand this to mean that the 
parameter $g$ does not encode small corrections to the variational ground 
state. This is unsurprising since $g$ is \textit{not} related a priori to 
the dimensionless ratio $J/h$. Rather, $g$ arises due to an artificial 
separation of the Hamiltonian into intra- and intercell terms. Therefore,
including higher-order-in-$g$ corrections will not necessarily improve the
estimate for the gap.

The main objective of this article is to explore Hirsch and Mazenko's 
renormalization-group perturbation method at second order in the real-space 
framework of Hamiltonian lattice gauge theory. To the best of our knowledge, 
a calculation based on such an approach has not been presented in the 
literature. The quantitative improvements for critical exponents garnered 
in this approach are modest compared to prior real-space findings---they are
not, by any means, state-of-the-art. We are able to
cross-check our calculations with Hirsch's for the transverse field
Ising model. The existence of a dual model without local symmetry is 
obviously helpful, but not requisite.

This article is organized as follows. In Section~\ref{sec:method} we review the
basic formalism of Fradkin and Raby's real-space renormalization-group approach
and how it fits into the perturbation theory of Hirsch and Mazenko. In 
Section~\ref{sec:provedecay} we pause to prove the exponential decay of spatial
correlations between $\mathbb{Z}_2$ magnetic monopoles in the ground state of
the electric free phase. Our results near and at criticality are presented in 
Section~\ref{sec:results}. A brief discussion is given in 
Section~\ref{sec:discuss}. The technical aspects of the renormalization
calculation are explained in detail in the appendix.

\section{Methodology \label{sec:method}}

\begin{figure}
\includegraphics{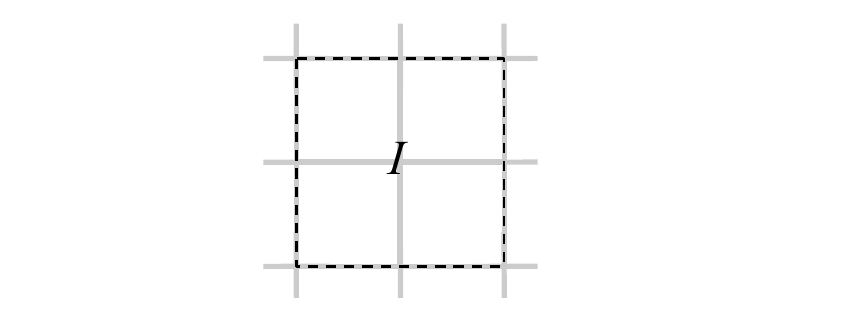}
\caption{\label{fig:blocking}
The cell $I$ consists of four plaquettes.}
\end{figure}

Following Fradkin and Raby, partition the lattice into regular, repeating 
square cells $I$ each comprised of four plaquettes $p$. See 
Fig.~\ref{fig:blocking}. A given link $l$ belonging to a cell is classified 
into one of two groups: internal (denoted by a dedicated index $i$)
corresponding to the four central links situated inside the cell, and external 
(denoted by a dedicated index $b$) corresponding to the eight boundary links 
around the edge of the cell. Let each cell have its own cellular Hamiltonian 
given by 
\begin{equation}
  H_I = -h\sum_{i\in I}\sigma^z_i - J\sum_{p\in I}\Phi_p.
\end{equation}
Since only the transverse field operators $\sigma^z_i$ of the four internal 
links are included, only these degrees of freedom act quantumly. The operators 
$\sigma^x_b$ of the eight external links behave classically and their 
eigenvalues serve as boundary conditions on the cell spectrum. We denote 
external link eigenstates and eigenvalues as
\begin{equation}
  \sigma^x_b\ket{x_b} = x_b\ket{x_b}, \quad x_b = \pm 1.
\end{equation}
Since cell $I$ contains four qubits and eight bits, and the cell Hamiltonian 
commutes with a generator of $\mathbb{Z}_2$ gauge transformations located at 
its center site, the cell Hilbert space has dimensionality $2^{4-1} = 8$ per 
boundary configuration. The spectrum is easily worked out analytically. Cell 
eigenstates depend parametrically on the $x_b$ around the boundary of the cell.
Let us denote the cell ground state as $\ket{0(\{x_b\})}_I$. Gauge symmetry 
constrains its energy eigenvalue $\epsilon^0_I$ to depend only on the product 
$\Phi_I = \prod_{b \in I}x_b$.

Define interactions by
\begin{equation}
  V = -h\sum_I\sum_{b\in I}\sigma^z_b,
\end{equation}
where it is understood that links are not repeated in the sum. The original 
lattice Hamiltonian is then
\begin{equation}\label{separateH}
  H = \sum_I H_I + g V,
\end{equation}
where the intercell coupling $g$ has been introduced to aid in organizing a 
perturbation expansion---its value is ultimately set to 1. 

The goal is to construct a renormalized Hamiltonian $H^\text{ren}$ governing a 
new set of spin-$\half$ operators $\{X_B,\,Z_B\}$ again obeying the Pauli 
algebra and defined on links $B$ corresponding to the sides of the cells $I$.
The renormalized electric and magnetic flux operators $Z_B$ and 
$\prod_{B\in I}X_B$ will then come with renormalized couplings $h'$ and $J'$, 
respectively. But we also expect that more complicated gauge-invariant 
operators are generated. This will proliferate more couplings. $H^\text{ren}$ 
is constructed such that, for an arbitrary configuration of the external link 
eigenvalues, its lowest eigenvalue is identical to that of $H$. At $g = 0$ this
is done by projecting $H$ onto a subspace spanned by states $\ket{n}$ which are
formed from tensor products over all cells $I$ of $\ket{0(\{x_b\})}_I$ and the 
$\ket{x_b}$ (without repetition). Such states have energy 
$\epsilon_n = \sum_I\epsilon^0_I(\Phi_I)$. The truncated Hilbert space is 
spanned by states $\ket{\mu_n}$ which are simply products of the $\ket{x_b}$. 
In essence, the internal links are decimated by the truncation. At the 
surviving links we define new operators 
$\mu^x_b = (\ket{+}\bra{+}-\ket{-}\bra{-})_b$ and 
$\mu^z_b = (\ket{+}\bra{-}+\ket{-}\bra{+})_b$. Then, for each pair of contiguous
links $b$ and $b'$ in the lattice, we define renormalized operators 
$X_B = \mu_b^x\mu_{b'}^x$ and $Z_B = (\mu_b^z+\mu_{b'}^z)/2$. 

There is a set of vectors $\{\ket{\alpha}\}$ much larger than and orthogonal to
the set $\{\ket{n}\}$ that, when combined with $\{\ket{n}\}$, span the original
Hilbert space. We construct $\ket{\alpha}$ similarly to $\ket{n}$ except that 
one or more cell eigenstates must be chosen in an excited cell state. For 
$g > 0$ the lowest eigenvalue of $H$ and $\sum_I H_I$ are not the same. 
Correcting this order by order in $g$ constrains the renormalized Hamiltonian 
to have the form \cite{HM}
\begin{subequations}\label{HirschMazenkoHren}
\begin{eqnarray}
  H^\text{ren} &=& H^{(0)} + g H^{(1)} + g^2 H^{(2)} + \dotsb, \\
  H^{(0)} &=& \sum_n \epsilon_n \ket{\mu_n}\bra{\mu_n} \\
  H^{(1)} &=& \sum_{n,n'} \ev{n'|V_\sigma|n}\ket{\mu_n}\bra{\mu_{n'}} \\
  H^{(2)} &=& \half\sum_{n,n'}\sum_\alpha
  \ev{n'|V_\sigma|\alpha}\ev{\alpha|V_\sigma|n} \nonumber \\
  & & \times \Bigl(\frac{1}{\epsilon_n-\epsilon_\alpha}+
  \frac{1}{\epsilon_{n'}-\epsilon_\alpha}\Bigr)
  \ket{\mu_n}\bra{\mu_{n'}}.
\end{eqnarray}
\end{subequations}
The detailed computation of these terms may be found in the appendix.

\section{Proof of exponential decay \label{sec:provedecay}}

\begin{figure}
\includegraphics{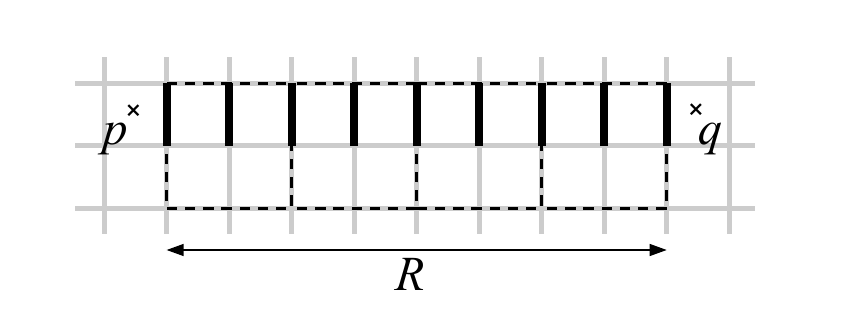}
\caption{\label{fig:correlator}
A correlation function of disorder operators at plaquettes at $p$ and $q$
separated by a line $\gamma$ of $R+1$ vertical links (bold). Attached to each
one of these links $l$ is a transverse field operator $\sigma^z_l$. One 
iteration of the decimation procedure (cells are shown dashed) reduces the 
separation to $R/2$. Links on the edges of the cells are boundary links and
operators living on these links do not act directly on the Hilbert space of the
cell.}
\end{figure}

Let $\lambda = J/h$. The electric free phase---the ground state in which lines
of electric flux can meander throughout the lattice without energy 
cost---corresponds to $\lambda \gg 1$. Denote the ground state of the lattice 
Hamiltonian by $\ket{\text{gs}_\lambda}$. 't Hooft disorder operators---which 
are string-like yet still local---create and annihilate magnetic monopoles. 
Their correlation function, when separated by a row of $R = 2^n$ plaquettes, 
is given by 
\begin{equation}\label{correlator}
  C_\lambda(R) = 
  \ev{\text{gs}_\lambda|\prod_{l\in\gamma}\sigma^z_l|\text{gs}_\lambda},
\end{equation}
where $\gamma$ is the set of $R+1$ vertical links separating the two plaquettes.
See Fig.~\ref{fig:correlator}. Fradkin and Raby's decimation procedure amounts 
to the approximate replacement
\begin{equation}
  \ket{\text{gs}_\lambda} \simeq 
  \prod_I\ket{0(\{x_b\})}_I \ket{\text{gs}_{\lambda'}}.
\end{equation}
Here $\ket{\text{gs}_{\lambda'}}$ is the ground state of $H^\text{ren}$ with 
$\lambda' = J'/h'$, which now has a fourth as many plaquettes. This is a 
truncation of the the original Hilbert space to a subspace of states that may 
be expressed solely in terms of boundary link eigenstates $\ket{x_b}$. 
Substitution gives
\begin{equation}
  C_\lambda(R) = 
  \bra{\text{gs}_{\lambda'}}\prod_b \sigma^z_b
  \prod_{I=1}^{R/2}\ev{0|\sigma^z_i|0}_I
  \ket{\text{gs}_{\lambda'}}.
\end{equation}
The appendix contains the explicit cell ground state wavefunction, 
Eq.~(\ref{pluswavefn0}) or Eq.~(\ref{minuswavefn}), and representation of the 
internal $\sigma^z_i$ matrices, Eq.~(\ref{internalZ}), needed to compute the
cell matrix element. Irrespective of the choice for $i$, the matrix element 
turns out to depend only on the sign of the magnetic flux, 
\begin{subequations}
\begin{equation}
  \ev{0|\sigma^z_i|0}_I = 
  \half(A_\lambda^- + A_\lambda^+)\Id_I 
  - \half(A_\lambda^- - A_\lambda^+)\Phi_I,
\end{equation}
where
\begin{eqnarray}
  A_\lambda^+ &=& 
  \frac{(1-\lambda^2+(1+\lambda^4)^{1/2})
  (1+\lambda^2+(1+\lambda^4)^{1/2})^{1/2}}{2\sqrt{2}(1+\lambda^4)^{1/2}}, \\
  A_\lambda^- &=& 
  \frac{1}{2}\Bigl(1+\frac{1}{(1+\lambda^2)^{1/2}}\Bigr).
\end{eqnarray}
\end{subequations}
To $O(g^1)$, the renormalized couplings $J'$ and $h'$ may be read off from 
Eqs.~(\ref{Jprime}) and (\ref{hprime}), respectively. For an initial choice of 
$\lambda \gg 1$, there is the asymptotic equivalence 
\begin{equation}
  \lambda' \approx 2\lambda.
\end{equation}
Since the renormalized coupling increases with iteration, the ground state has 
no magnetic energy and the operator $\Phi_I$ evaluates to $+1$. In terms of
renormalized link operators, this yields the multiplicative recursion relation
\begin{equation}
  C_\lambda(R) \approx (A^+_\lambda)^{R/2}C_{\lambda'}(R/2).
\end{equation}
Iterating $n$ times starting from $\lambda_0$,
\begin{equation}
  C_{\lambda_0}(2^n) = 
  (A^+_{\lambda_0})^{2^{n-1}}
  (A^+_{\lambda_1})^{2^{n-2}}
  \dotsb
  (A^+_{\lambda_{n-1}})^{2^{n-n}}
  C_{\lambda_n}(1).
\end{equation}
Substituting $A^+_\lambda \approx 1/2\lambda$ and $\lambda_n \approx 
2^n\lambda_0$ yields
\begin{equation}
  C_{\lambda_0}(2^n) \approx 
  \frac{2^n}{(4\lambda_0)^{2^n-1}}C_{2^n\lambda_0}(1).
\end{equation}
But $C_\lambda(1) \sim \lambda^{-1}$ up to some constant factor, so 
\begin{equation}
  C_{\lambda_0}(R) \sim 
  \frac{4R}{(4\lambda_0)^R} \propto \exp(-R\log(4\lambda_0)).
\end{equation}

\section{Results \label{sec:results}}

\begin{figure}
\includegraphics{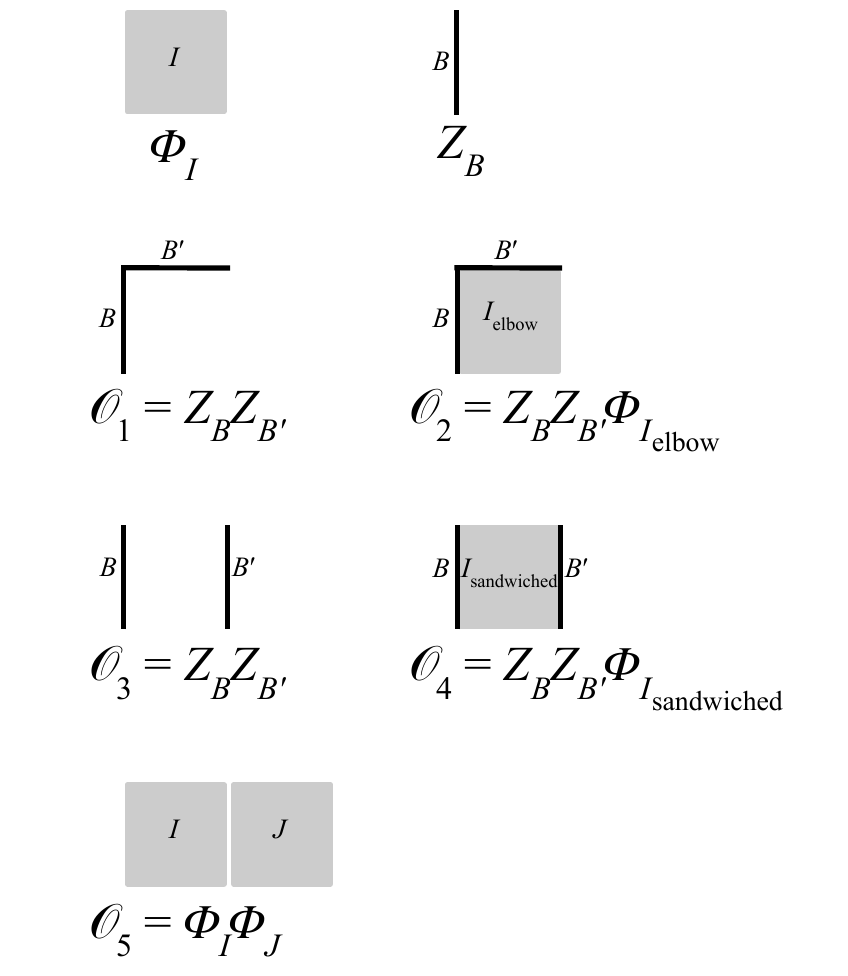}
\caption{\label{fig:operators}
Effective operators present in renormalized Hamiltonian $H^\text{ren}$ to 
second order in the intercell coupling $g$. These are dual to the operators
found by Hirsch in his study of the two-dimensional Ising model in a transverse
field. \cite{Hirsch}}
\end{figure}

Our main technical achievement is the explicit expression for the renormalized
Hamiltonian calculated from Eq.~(\ref{HirschMazenkoHren}). The details of the
calculation are given in the appendix. The reader interested only in the final
form of $H^\text{ren}$ can see the precise operators in Eq.~(\ref{recur}), 
although several definitions needed to understand the coefficients of these 
operators are scattered throughout the appendix. It turns out that five new 
effective operators are created in addition to the effective electric flux 
$h Z_B$ on each link $B$, and effective magnetic flux $J\Phi_I$ on each cell 
$I$. We denote these new operators as $K_\alpha\mathcal{O}_\alpha$ for 
$\alpha = 1,\dotsc,5$. See Fig.~\ref{fig:operators}. Also present is the 
identity $F\Id_I$ on each cell $I$. 

\subsection{Critical point}

\begin{figure}
\includegraphics{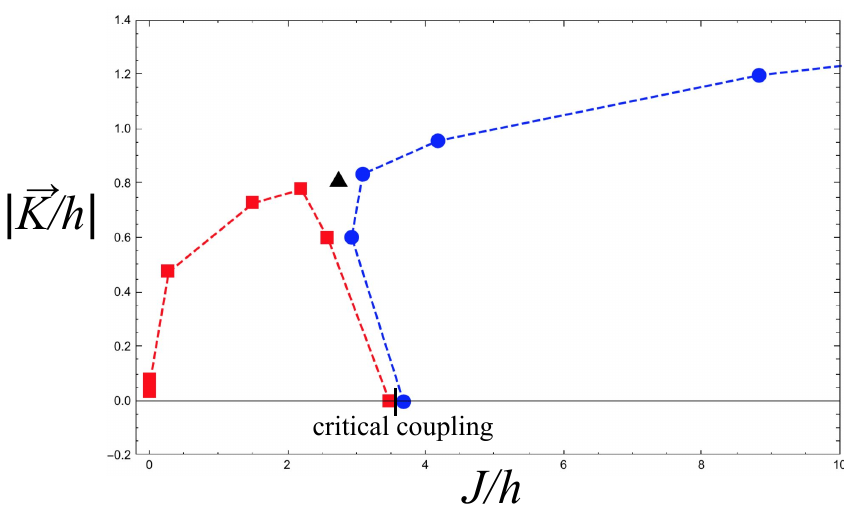}
\caption{\label{fig:discreteflow}
Plot of discrete RG flows beginning from the $\vec{K}/h = 0$ axis 
projected into the $(J/h,|\vec{K}/h|)$ plane (color online). 
The nontrivial fixed point is indicated by a black triangle. 
An initial coupling just above the critical value initiates 
a flow in which $J/h$ tends
to grow without bound (blue dots). An initial coupling just below the critical
value produces a flow that terminates near the origin (red squares). The
critical surface lies in-between these flows, and connects the critical coupling
to the nontrivial fixed point. 
}
\end{figure}

In the appendix we study numerically the recursion relations for the operator
coefficients $h$, $J$, $K_\alpha$, and $F$. Following Fradkin and Raby, we
adopt $h$ as an energy scale and consider the dimensionless couplings 
$(J/h,K_\alpha/h)$ packaged into a six-dimensional vector. Iterations of the 
recursion relations produce a sequence of points in this vector space (a 
discrete ``RG flow'') 
that describes Hamiltonians defined over successively coarser lattices. 
We are interested in flows that begin on the axis $(J/h,\vec{0})$. 
Our numerical analysis was implemented in {\it Mathematica 10}.
One way to visualize the flow is shown in Fig.~\ref{fig:discreteflow}, where
the sequence of points in the full six-dimensional space has been projected 
down to the plane $(J/h,|\vec{K}/h|)$. On the line given by $\vec{K}/h = 0$,
there exists a critical coupling $(J/h)_\text{c} = 3.56895$ for which the flow 
converges onto a nontrivial fixed point $(J/h,\vec{K}/h)_*$. This fixed point,
whose full coordinates are given in the appendix, 
was located using Newton's root-finding method 
yielding very fast convergence in only several iterations 
starting from the initial point $(J/h,\vec{K}/h) = (3.3,\vec{0})$. Our 
fine-tuning estimate for the critical coupling was obtained by searching for 
the flow that spent the greatest number of steps near the fixed point within
some tolerance.

Generically, for $J/h \neq (J/h)_\text{c}$, flows eventually veer away from 
the fixed point and tend toward either the origin or grow unbounded as suggested
in Fig.~\ref{fig:discreteflow}. Thus, the nontrivial
fixed point is unstable and infrared-repulsive, whilst the trivial fixed points
at the origin and infinity are stable and infrared-attractive. 
We note that couplings in the electric free phase for
$J/h$ slightly larger than $(J/h)_\text{c}$ give rise to flows in which $J/h$ 
gets extremely large relative to $|\vec{K}/h|$, and while 
there appears to be no bound
on this growth, the sign of $J/h$ eventually alternates which calls into
question the asymptotic reliability of the recursion relation.

In the 
neighborhood of the nontrivial fixed point there is a linear space with a 
single relevant scaling variable that we call $u_1$. Iterations of the recursion
relations renormalize $u_1$ to $\Lambda_1 u_1$, where $\Lambda_1 > 1$ is an
eigenvalue computed in the appendix.

\subsection{Energy gap}

Consider the electric free phase in which $J/h$ is only slightly greater than 
$(J/h)_\text{c}$. The flow is observed to behave as 
follows: a small and finite number of steps $N_1$ brings the flow 
from $(J/h,\vec{0})$ into the neighborhood of the nontrivial fixed point; for
some large, but finite, number of steps $N_2$, the flow dawdles and remains
quite close to the fixed point; further steps finally allow the flow to escape
this region and head off to infinity. Numerical analysis shows that whilst the
five couplings $K_\alpha/h$ remain of order one, the coupling $J/h$ grows
without bound. This is expected since the trivial fixed point at infinity should
describe the deconfined phase of the lattice gauge theory with infinitely 
heavy magnetic monopoles. So now consider the recursion relations just for the
coefficients $h$ and $J$, which may be expressed in the form
\begin{subequations}
\begin{eqnarray}
  h' & = & h\zeta(J/h,\vec{K}/h), \\
  J' & = & J\eta(J/h,\vec{K}/h).
\end{eqnarray}
\end{subequations}
Specifically, the function $\zeta$ is given by dividing the right-hand side of 
Eq.~(\ref{hrecur}) by $h$, and the function $\eta$ is given by dividing the
right-hand side of Eq.~(\ref{Jrecur}) by $J$. Numerical evidence suggests that
repeated iterations cause $\zeta$ to approach some number less than 1, and
$\eta$ to approach 1. \footnote{At order $g^1$, it is easy to show analytically
that $\lim_{J/h \to \infty}\zeta = \half$ and $\lim_{J/h \to \infty}\eta = 1$.}
Therefore, in the limit of infinitely many renormalization-group iterations
the coefficient $h$ will vanish and only $J$ will remain. Thus, the original
Hamiltonian defined on the fine lattice will exhibit the same energy gap as a
Hamiltonian defined on the coarse lattice containing only the operators 
$\sum_I\Phi_I$. Since the latter theory is weakly coupled, it may be analyzed 
perturbatively.

The ground state is characterized by the absence of magnetic flux for all 
plaquettes (i.e., $\Phi_I = +1$ for all $I$). The lowest energy excitation 
creates a unit of magnetic flux on a single plaquette. The energy of the first
excited state (relative to the ground state) is $G \simeq 2J_N$, where $N$ 
indicates the total number of iterations of the recursion relations. Since 
$J_N = J_0\prod_{n=0}^{N-1}\eta(J_n/h_n,\vec{K}_n/h_n)$, we need only keep a 
record of the point sequence along the flow in order to calculate the gap. 
However, close to criticality the product may be decomposed as
\begin{equation}\label{gap}
  G \simeq 2J_0
  \prod_{n=0}^{N_1-1}\eta 
  \prod_{n=N_1}^{N_1+N_2}\eta
  \prod_{n=N_1+N_2}^N\eta.
\end{equation}
The first product in Eq.~(\ref{gap}) corresponds to the inflow. It will be 
analytic in the difference $J/h - (J/h)_\text{c}$. The last product corresponds
to the outflow and therefore we expect it asymptotes to the value 1. However,
the middle product corresponds to the dawdle near the fixed point. We can 
evaluate $\eta$ at the fixed point---this approximation gets better the closer
the flow starts to criticality. The number $N_2$ may be estimated by asking how
many steps need to be taken to multiplicatively renormalize the relevant 
scaling variable $u_1$---which we assume is exceedingly small at step 
$N_1$---into an arbitrary, but fixed and small number $U_1$ for which the 
linearized approximation to the recursion relations is still valid. This
condition is
\begin{equation}\label{estimateN}
  U_1 = \Lambda_1^{N_2}u_1 \implies N_2 = \log(U_1/u_1)/\log \Lambda_1.
\end{equation}
Hence,
\begin{equation}
  \text{nonanalytic part of $G$} \sim \eta((J/h,\vec{K}/h)_*)^{N_2}.
\end{equation}
Since $u_1$ arises from the inflow and $N_1$ is finite, it follows that $u_1$ 
itself is some analytic function of $J/h - (J/h)_\text{c}$. Finally,
\begin{equation}
  \text{nonanalytic part of $G$} \sim 
  (J/h - (J/h)_\text{c})^{\nu_t},
\end{equation}
where
\begin{equation}
  \nu_t = \frac{-\log\eta((J/h,\vec{K}/h)_*)}{\log\Lambda_1} \simeq 0.49.
\end{equation}
Obtaining this critical exponent, which was not computed in 
Ref.~\onlinecite{Hirsch},
was one of the original motivations for this work.

\subsection{Spatial correlation length}

In the electric free phase, the correlation function of disorder operators
given by Eq.~(\ref{correlator}) ought to have a correlation length $\xi$
that diverges as $J/h$ approaches $(J/h)_\text{c}$ from above. After some large 
number $N$ of decimations, the dimensionless correlation length will be an
order one number because the flow will be far from the neighborhood of the 
fixed point where the linearized recursion relations hold. Therefore, the part
of $N$ that depends on the reduced coupling $J/h-(J/h)_\text{c}$ must be the 
same as $N_2$ as estimated in Eq.~(\ref{estimateN}). Since our renormalization
scale factor is $2$,
\begin{equation}
  \xi \sim 2^N \sim (J/h-(J/h)_\text{c})^{\nu_s},
\end{equation}
where
\begin{equation}
  \nu_s = \frac{\log 2}{\log \Lambda_1} \simeq 0.65.
\end{equation}

\subsection{Ground state energy}

From the coefficient of the identity operator on each cell it is possible to
represent the ground state energy by the expression
\begin{equation}
  E_\text{gs}(J/h,\vec{K}/h) = \lim_{n\to\infty} F^{(n)}\frac{N_\text{plaq}}{
  4^n},
\end{equation}
where the superscript $^{(n)}$ denotes the $n$th iteration of the recursion
relation given by Eq.~(\ref{Frecur}). The case $n = 0$ indicates the original
bare coupling. For instance, $h^{(0)} = h$. Define the energy density by 
$\varepsilon_\text{gs} = E_\text{gs}/N_\text{plaq}$. Rather than a limit, let 
us express the ground state energy density as an infinite sum. For clarity, we 
write the six-dimensional vector of dimensionless couplings as 
$\kappa^{(n)} = (J^{(n)}/h^{(n)}, \vec{K}^{(n)}/h^{(n)})$. Knowing that the
recursion relation for $F$, Eq.~(\ref{Frecur}), takes the form \footnote{
Specifically, $\Delta$ is everything on the right-hand side of 
Eq.~(\ref{Frecur}) with $4F$ set to $0$ and $h$ set to $1$.}
\begin{equation}
  F^{(n+1)} = 4F^{(n)} + h^{(n)}\Delta(\kappa^{(n)}),
\end{equation}
where $\Delta$ is an analytic function of its argument, we obtain 
\begin{equation}\label{egssum}
  \varepsilon_\text{gs}(\kappa^{(0)}) = \sum_{n = 0}^\infty 
  \frac{1}{4^{n+1}}h^{(n)}\Delta(\kappa^{(n)}).
\end{equation}
We have assumed that $F^{(0)} = 0$. Since the recursion relation for $h$, 
Eq.~(\ref{hrecur}), takes the form
\begin{equation}
  h^{(n+1)} = h^{(n)}\zeta(\kappa^{(n)}),
\end{equation}
it follows that
\begin{equation}
  h^{(n)} = h^{(0)}\zeta(\kappa^{(0)})\zeta(\kappa^{(1)})\dotsb
  \zeta(\kappa^{(n-1)}).
\end{equation}
Therefore, in Eq.~(\ref{egssum}), specification of $\kappa^{(0)}$ completely
determines the right-hand side since the recursion relations may be applied to
to obtain $\kappa^{(1)}$, $\kappa^{(2)}$, etc. Following Ref.~\onlinecite{MG}, 
Eq.~(\ref{egssum}) may be written
\begin{widetext}
\begin{equation}\label{egssumlong}
  \varepsilon_\text{gs}(\kappa^{(0)}) =  
  \tfrac{1}{4}h^{(0)}\Delta(\kappa^{(0)}) + 
  \tfrac{1}{4}\zeta(\kappa^{(0)})\biggl[
  \tfrac{1}{4}h^{(0)}\Delta(\kappa^{(1)}) + 
  \tfrac{1}{4^2}h^{(0)}\zeta(\kappa^{(1)})\Delta(\kappa^{(2)}) +
  \tfrac{1}{4^3}h^{(0)}\zeta(\kappa^{(1)})\zeta(\kappa^{(2)})
  \Delta(\kappa^{(3)}) + \dotsb\biggr].
\end{equation}
\end{widetext}
Notice that the bracketed term in Eq.~(\ref{egssumlong}) is just the right-hand
side of Eq.~(\ref{egssum}) but started at the point $\kappa^{(1)}$. Thus, we 
arrive at a recursion relation satisfied by the ground state energy density,
\cite{MG}
\begin{equation}\label{egsrecur}
  \varepsilon_\text{gs}(\kappa^{(0)}) = \tfrac{1}{4}h^{(0)}\Delta(\kappa^{(0)})
  + \tfrac{1}{4}\zeta(\kappa^{(0)})\varepsilon_\text{gs}(\kappa^{(1)}).
\end{equation}

We are interested in extracting the leading singular behavior of the ground
state energy density as the critical point is approached. Since $\Delta$ is 
differentiable, even at the fixed point, Eq.~(\ref{egsrecur}) implies the 
following homogeneous transformation law for the singular part of 
$\varepsilon_\text{gs}$,
\begin{equation}
  \varepsilon_\text{gs}^\text{sing}(\kappa) = 
  \frac{1}{4}\zeta(\kappa)\varepsilon_\text{gs}^\text{sing}(\kappa').
\end{equation}
Close to the fixed point, we can write this using scaling variables. Ignoring
irrelevant variables and iterating $n$ times,
\begin{equation}
  \varepsilon_\text{gs}^\text{sing}(u_1) = 4^{-n}
  \prod_{r=0}^{n-1}\zeta(\Lambda_1^r u_1)
  \varepsilon_\text{gs}^\text{sing}(\Lambda_1^n u_1).
\end{equation}
Since $u_1$ grows under iteration we need to apply a stopping condition. We take
$n = N_2$ as specified by Eq.~(\ref{estimateN}). If $\Lambda_1^r u_1$ remains
small, then we may approximate each $\zeta$ by its value at the fixed point
$u_1 = 0$. Then
\begin{equation}
  \varepsilon_\text{gs}^\text{sing}(u_1) \approx (4/\zeta(0))^{-n}
  \varepsilon_\text{gs}^\text{sing}(U_1) \sim 
  u_1^{\frac{\log(4/\zeta(0))}{\log \Lambda_1}}. 
\end{equation}
Finally,
\begin{equation}
  \text{singular part of $\varepsilon_\text{gs}$} \sim 
  |J/h - (J/h)_\text{c}|^{2-\alpha},
\end{equation}
where
\begin{equation}\label{alphaexp}
  \alpha = 2 - \frac{\log(4/\zeta(0))}{\log \Lambda_1} \simeq 0.21. 
\end{equation}

A direct numerical calculation of Eq.~(\ref{egssum}) at the critical coupling 
$(J/h)_\text{c}$ produces $\varepsilon_\text{gs} \simeq -3.718h^{(0)}$.

\subsection{A critical amplitude ratio}

\begin{figure}
\includegraphics{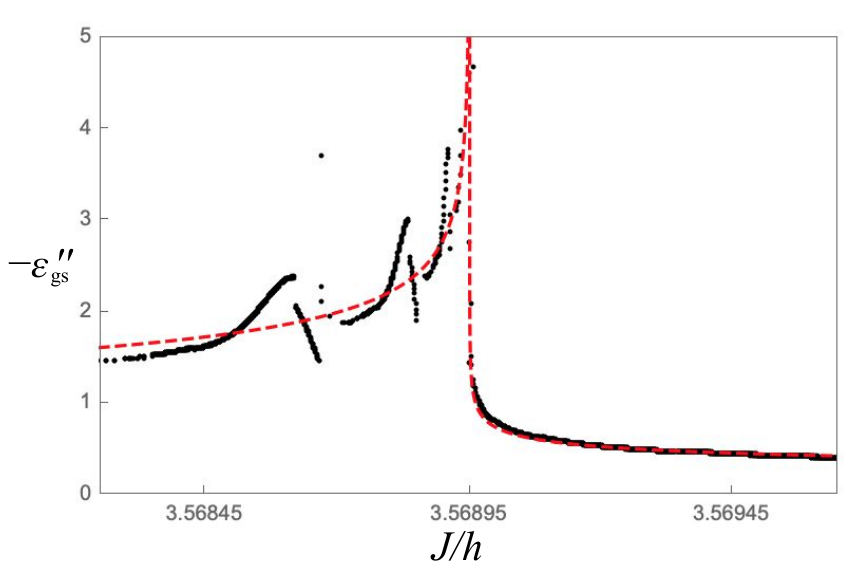}
\caption{\label{fig:specificheat}
A plot of a numerical calculation of 
$-\partial^2\varepsilon_\text{gs}/\partial(J/h)^2$. It diverges at
the critical coupling $(J/h)_\text{c} = 3.56895$. Fits to the leading
singular behavior are shown as dashed red lines (color online).
The numerous local peaks and valleys below the critical coupling 
are spurious artifacts of our numerical method and should be ignored.}
\end{figure}

We also studied the divergence of the second derivative of the ground state
energy density with respect to $J/h$, denoted $\varepsilon_\text{gs}''$, 
in a small region around the critical coupling. A plot of this 
divergence is visible in Fig.~\ref{fig:specificheat} as the prominent
spike. The main difficulty in obtaining these values, besides discretization
error associated with differentiating, is the inability to evaluate 
Eq.~(\ref{egssum}) to arbitrarily high $n$. Our renormalization group 
transformation $\kappa^{(n)} \to \kappa^{(n+1)}$ cannot be iterated 
indefinitely without running into a nonsensical value for $J/h$ (e.g., negative
values). In practice, we could iterate the flow between 10 and 20 steps, more
steps being possible the closer $J/h$ begins to the true critical coupling.

The leading singular behavior in $\varepsilon_\text{gs}$ comes from that part
of the sum in Eq.~(\ref{egssum}) corresponding to the renormalization flow along
the outflow trajectory (see Fig.~\ref{fig:flow}) \cite{Cardy}. Since the flow
away from the fixed point is different for the two phases of the gauge theory,
we expect the amplitude $A$ to be different for $J/h > (J/h)_\text{c}$ and 
$J/h < (J/h)_\text{c}$. We may write
\begin{equation}
  \text{leading singular part of $\varepsilon_\text{gs}$} \sim 
  A_{>,<}|J/h - (J/h)_\text{c}|^{2-\alpha}.
\end{equation}
The amplitude ratio $A_>/A_<$ is a universal quantity, but unlike critical
exponents, it depends on the entire flow, not just the linearized flow in the
vicinity of the fixed point.

We applied a naive procedure to obtain a cursory estimate for the amplitudes. 
After transforming data to the form $(\log|J/h-(J/h)_\text{c}|,\,
\log\varepsilon_\text{gs}'')$, we made a least-squares fit to the line 
$y = b_0 - \alpha x$, where $b_0$ is the single free parameter and $\alpha$ is 
constrained to be the value in Eq.~(\ref{alphaexp}). We remark that, even with 
a two-parameter fit like $y = b_0 - b_1 x$, the slope parameter $b_1$ comes 
within $10\%$ of $\alpha$. Notice that we completely ignore 
correction-to-scaling terms in these fits. 
We obtain $b_0^> = -2.368$ and $b_0^< = -1.036$.
Then
\begin{equation}
  A_>/A_< = \exp(b_0^> - b_0^<) \simeq 0.26.
\end{equation}
For comparison, the ratio of specific heat amplitudes in the three-dimensional
Ising universality class is known to be about 0.52. \cite{Privman}

\begin{table}
\begin{ruledtabular}
\begin{tabular}{ccc}
  order in Hirsch--Mazenko & & \\
  perturbation theory & $(J/h)_\text{c}$ & $E_\text{gs}/hN_\text{plaq}$ \\
  \hline
  first (Refs.~\onlinecite{FR,MG}) & 3.28 & $-3.376$ \\
  second (Ref.~\onlinecite{Hirsch} and this work) & 3.57 & $-3.718$
\end{tabular}
\end{ruledtabular}
\caption{\label{tab:nonuniver}
Critical coupling $(J/h)_\text{c}$ and critical ground state energy per
plaquette $E_\text{gs}/hN_\text{plaq}$ 
in the Hirsch--Mazenko perturbation expansion. 
}
\end{table}

\section{Discussion \label{sec:discuss}}

Using Hirsch--Mazenko perturbation theory we have calculated some critical 
properties of the quantum $\mathbb{Z}_2$ gauge theory on a square 
lattice. Universal critical exponents are given in 
Table~\ref{tab:critexp} while nonuniversal data are collected in 
Table~\ref{tab:nonuniver}. Most indications are that the second-order 
theory is an improvement over the first-order theory.

It is known from Monte Carlo simulations of the simple cubic Ising model
that the critical inverse temperature is $K_\text{c} \simeq 0.22$ 
and the critical exponent for the correlation length is 
$\nu_\text{Ising} \simeq 0.63$ \cite{FXL}. 
Hyperscaling then implies that the critical exponent for the specific
heat is $\alpha_\text{Ising} = 2-3\nu_\text{Ising} \simeq 0.11$.
Using duality we are able to transfer these values over to the gauge theory:
the critical coupling ought to be $(J/h)_\text{c} = K_\text{c}^{-1} \simeq 
4.51$, the critical exponents for the energy gap and spatial correlation 
length---equal due to rotational invariance at the critical point---ought to be
$\nu_s = \nu_t = \nu_\text{Ising} \simeq 0.63$, and the critical exponent for
the leading singular behavior of the ground state energy ought to be
$\alpha = \alpha_\text{Ising} \simeq 0.11$. 

Most of our second-order results
are closer to these expected values than the first-order results. Of particular
note is that $\nu_s$ and $\alpha$, which we stress were computed independently,
both improved dramatically at second order. To wit, $\alpha$ made a
qualitative switch from negative to positive!

Furthermore, at the fixed point, we find that the reciprocal of the gap 
energy scales, under a renormalization transformation, 
by a factor of $1/\eta((J/h,\vec{K}/h)_*) \simeq 1.68$. This is 
certainly closer to the spatial scale factor of $2$ than the $O(g^1)$ result of
$1.43$. This supports Fradkin and Raby's suggestion that systematic improvement
is possible using a perturbative framework like Hirsch and Mazenko's.

In order to evaluate the accuracy of the critical ground state energy density,
we may use a duality relation between quantum Hamiltonians for the
two-dimensional $\mathbb{Z}_2$ lattice gauge theory (``LGT'') and the 
two-dimensional transverse field Ising model (``TFIM'') \cite{Kogut}. If
$\lambda = J/h$ and $E$ is any eigenvalue, then 
$E_\text{LGT}(\lambda) = \lambda E_\text{TFIM}(\lambda^{-1})$. Using the 
critical data in Table~\ref{tab:nonuniver}, the critical ground state energy 
per spin in the TFIM is approximately $-1.03$ (first order) and $-1.04$ (second
order). A numerical calculation of the lowest eigenvalue of the TFIM 
Hamiltonian on a $4 \times 4$ lattice evaluated at coupling $K_c$ yields a
ground state energy per spin of $-1.02$. The agreement is decent. 

Unfortunately, not all Hirsch--Mazenko perturbative corrections are 
improvements. There is a clear worsening of the gap critical exponent: the
second-order value for $\nu_t$ is much 
worse than its first-order value. This suggests that the artificial 
separation of the Hamiltonian given by Eq.~(\ref{separateH}) is not a small
correction to the variational ground state energy.

\begin{table}
\begin{ruledtabular}
\begin{tabular}{lccc}
  cell size & $(J/h)_\text{c}$ & $\nu_t$ & $\nu_s$ \\
  \hline
  $2\times 2$ & 3.280 & 0.622 & 1.197 \\
  $3\times 3$ & 3.036 & 0.624 & 1.006 \\ 
  $4\times 4$ & 2.970 & 0.627 & 0.924
\end{tabular}
\end{ruledtabular}
\caption{\label{tab:cellsize}
Critical coupling and exponents at $O(g^1)$ in the Hirsch--Mazenko 
perturbation expansion for different cell sizes. These are most easily 
computed from the transverse field Ising model via duality. \cite{DWY,Sol}
}
\end{table}

One may straightforwardly improve $\nu_t$ by enlarging the cell size. For 
instance, we have done an $O(g^1)$ analysis using $3\times 3$ and $4\times 4$
cells. Our 
results are summarized in Table~\ref{tab:cellsize}. There is modest improvement after increasing the cell Hilbert space dimensionality from $2^4$ to $2^9$ and
then to $2^{16}$, 
indicating that the variational approximation is a little better.

What lessons and future directions does our work suggest? We have confirmed 
Fradkin and Raby's speculation that Hirsch--Mazenko perturbation theory can be
applied to a lattice gauge theory and that a second-order correction of their
renormalization transformation does lead to some qualitative improvements in 
the critical behavior. However, it is clear that precisely approximating 
critical exponents is not a strength of the real-space method. The
attractiveness of the method is tempered by the fact that some critical 
exponents (e.g., $\nu_t$) can get worse.
Therefore, the most promising and fruitful use of this work would be to guide
explorations of more complicated gauge-invariant Hamiltonians in two dimensions.
Preserving gauge invariance at successive steps in the 
renormalization process is nontrivial, but we have demonstrated explicitly
how it happens for the simplest gauge group.


\appendix*

\section{Details}

In this appendix we use a different and more thorough notation than in the main
body of the article.

\subsection{Notation}

\begin{figure}
\includegraphics{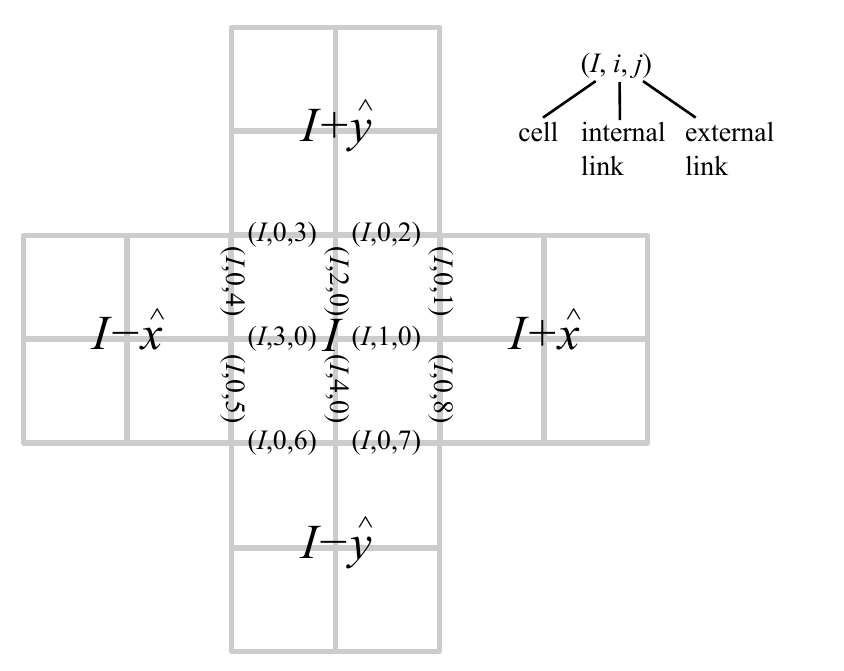}
\caption{\label{fig:cellnotate}
Our notation for the links.}
\end{figure}

Let $I$ denote a cell of four plaquettes. See Fig.~\ref{fig:cellnotate}. The 
four neighboring cells are called $I + \xhat$, $I - \xhat$, $I + \yhat$, and
$I - \yhat$. Each cell has four internal links with spin-$\half$ operators
$\vec{\sigma}_{I,i,0}$, $i=1,\dotsc,4$. Also, each cell is surrounded by eight 
external links with spin-$\half$ operators $\vec{\sigma}_{I,0,j}$, 
$j=1,\dotsc,8$. The ordering convention is explained in the figure.

Since external links that are on the boundary of a given cell are shared in 
common with one other neighboring cell, it will be useful to have a notation 
denoting an equivalent link from the perspective of the neighbor. If $(I,0,j)$ 
is a given link in a given cell, then the neighboring cell that shares that 
link will be denoted $\Inn$, and the same link will, from this cell's 
perspective, be called $[j]$. Explicitly,
\begin{eqnarray}
  (I,0,j) & = & (\Inn,0,[j]) \label{bracketmap} \\
  (I,0,8) & = & (I+\xhat,0,5) \nonumber \\
  (I,0,1) & = & (I+\xhat,0,4) \nonumber \\
  (I,0,2) & = & (I+\yhat,0,7) \nonumber \\
  (I,0,3) & = & (I+\yhat,0,6) \nonumber \\
  (I,0,4) & = & (I-\xhat,0,1) \nonumber \\
  (I,0,5) & = & (I-\xhat,0,8) \nonumber \\
  (I,0,6) & = & (I-\yhat,0,3) \nonumber \\
  (I,0,7) & = & (I-\yhat,0,2). \nonumber
\end{eqnarray}

Separate the Hamiltonian into an intracell part and intercell part,
\begin{equation}
  \label{Hseparated}
  H_\sigma = H_\sigma^0 + V_\sigma.
\end{equation}
The intracell part is a sum over all cells of the internal links, including 
both $\sigma^x$ and $\sigma^z$ operators, and the external links, but including
only the $\sigma^x$ operators. Let
\begin{subequations}
\begin{equation}
  H_\sigma^0 = \sum_I H_I^0,
\end{equation}
where
\begin{equation}
  H_I^0 = -h\sum_{i=1}^4\sigma^z_{I,i,0} - J\sum_{i=1}^4
  \sigma^x_{I,i,0}\sigma^x_{I,i+1,0}\sigma^x_{I,0,2i-1}\sigma^x_{I,0,2i}.
\end{equation}
\end{subequations}
The intercell part is a collection of all transverse field operators acting on 
the external links of each cell,
\begin{equation}
  V_\sigma = -h\sum_I \sum_{j=1}^8\sigma^z_{I,0,j}.
\end{equation}
It is understood that all external links are to be summed over just once. As a 
shorthand we will write $\sum_{I,j}$.

It is important to notate eigenvalues of $\sigma^x$ operators on external links.
For any $I$ and $j$, denote
\begin{equation}
  \sigma^x_{I,0,j}\ket{x_{I,j}}_{I,0,j} = x_{I,j}\ket{x_{I,j}}_{I,0,j}, 
  \quad x_{I,j} = \pm 1.
\end{equation}
Since $[H_I^0,\sigma^x_{I,0,j}] = 0$ for each $j$, each external link operator 
$\sigma^x_{I,0,j}$ may be replaced by its eigenvalue $x_{I,j}$. Thus, the eight
bits $x_{I,j}$ behave as classical boundary conditions. For the cell 
Hamiltonian we may write $H_I^0(x_{I,1},\dotsc,x_{I,8})$.

Each cell, because it has four qubits and eight bits, would seem to have a 
$2^4$-dimensional Hilbert space for every classical configuration of its 
external links. However, gauge invariance reduces this large space of 
possibilities so that, ultimately, it matches the information encoded in the 
quantum Ising model with a block of four sites. First, 
$[H_I^0, \prod_{i=1}^4\sigma^z_{I,i,0}] = 0$ and we are interested only in the 
gauge-invariant sector $\prod_i \sigma^z_{I,i,0} = +1$. This halves the 
dimensionality from $2^4$ to $2^3$. And each eigenstate of $H_I^0$ has an 
eigenvalue that depends only on the sign of the gauge-invariant flux operator,
\begin{equation}
  \Phi_I = \prod_{j=1}^8\sigma^x_{I,0,j} = \prod_{j=1}^8 x_{I,j}.
\end{equation}
That is, if we denote an eigenstate of $H_I^0$ by
\begin{equation}
  \ket{i_I(x_{I,1},\dotsc,x_{I,8})}_I, \quad i_I = 0,\dotsc,7,
\end{equation}
then its corresponding eigenvalue is some
\begin{equation}
  \epsilon_{i_I}^\text{c}(\Phi_I).
\end{equation}
The superscript ``c'' stands for ``cell.'' Thus, we are really dealing with a 
Hilbert space containing 16 eigenstates: 8 in the sector with $\Phi_I = +1$, 
and 8 in the sector with $\Phi_I = -1$. We reserve $i_I = 0$ to indicate the 
lowest-energy state in either sector.

\subsection{Cell spectrum}

The explicit wavefunctions and energies for $H_I^0$ may be worked out by 
following the procedure in Ref.~\onlinecite{FR}. We use a basis for the 
internal links in which $\sigma^z_{I,i,0}$ is diagonal (i.e., 
$\sigma^z_{I,i,0}\!\up_i\,=\,\up_i$ and 
$\sigma^z_{I,i,0}\!\dn_i\,=-\!\dn_i$, and 
$\sigma^x_{I,i,0}\!\up_i\,=\,\dn_i$ and
$\sigma^x_{I,i,0}\!\dn_i\,=\,\up_i$). The basis ordering is 
$\{\up_1\up_2\up_3\up_4,\,\up\dn\up\dn,\,\dn\up\dn\up,\,\dn\dn\dn\dn,\,
\up\up\dn\dn,\,\dn\dn\up\up,\,\up\dn\dn\up,\,\dn\up\up\dn\}$. Since all 
eigenstates are going to be expressed in this basis, the condition 
$\prod_i \sigma^z_{I,i,0} = +1$ is automatically satisfied. Define
\begin{equation}
  A_k = x_{I,2k-1}x_{I,2k}, \quad k = 1,\dotsc,4.
\end{equation}
Then
\begin{subequations}
\begin{equation}
  H_I^0 = \begin{pmatrix} P & Q \\ Q^T & 0 \end{pmatrix},
\end{equation}
where
\begin{eqnarray}
  P &=& -4h
  \begin{pmatrix} 1 & & & \\ & 0 & & \\ & & 0 & \\ & & & -1 \end{pmatrix}, \\
  Q &=& -J\begin{pmatrix}
  A_3 & A_1 & A_2 & A_4 \\
  A_2 & A_4 & A_3 & A_1 \\
  A_4 & A_2 & A_1 & A_3 \\
  A_1 & A_3 & A_4 & A_2 \\
\end{pmatrix}.
\end{eqnarray}
\end{subequations}
Clearly, the wavefunctions are not functions of the eight $x_{I,j}$, but rather
the four combinations $A_k$,
\begin{equation}
  \ket{i_I(x_{I,1},\dotsc,x_{I,8})}_I = \ket{i_I(A_1,\dotsc,A_4)}_I.
\end{equation}

We wish to solve the eigenvalue problem $H_I^0\ket{i_I}_I = 
\epsilon_{i_I}^\text{c}\ket{i_I}_I$ subject to additional constraints 
inherited from gauge invariance. In the full Hamiltonian $H$, gauge 
transformations are possible at each of the nine sites in cell $I$. However, 
because of the artificial nature of the blocking scheme the eight 
transformations around the cell perimeter no longer manifest as symmetries from
the point of view of the cell Hamiltonian $H_I^0$. Instead, they manifest as 
the following identities (for the sake of brevity we only write parameters that
are being affected in some way, e.g., by being negated):
\begin{subequations}
\label{cornerrules}
\begin{eqnarray}
  H_I^0(x_{I,1},x_{I,2}) &=& H_I^0(-x_{I,1},-x_{I,2}), \\
  H_I^0(x_{I,3},x_{I,4}) &=& H_I^0(-x_{I,3},-x_{I,4}), \\
  H_I^0(x_{I,5},x_{I,6}) &=& H_I^0(-x_{I,5},-x_{I,6}), \\ 
  H_I^0(x_{I,7},x_{I,8}) &=& H_I^0(-x_{I,7},-x_{I,8}),
\end{eqnarray}
\end{subequations}
and
\begin{subequations}
\label{midptrules}
\begin{eqnarray}
  \sigma^z_{I,1,0}H_I^0(x_{I,8},x_{I,1})\sigma^z_{I,1,0} 
  &=& H_I^0(-x_{I,8},-x_{I,1}),\qquad \\
  \sigma^z_{I,2,0}H_I^0(x_{I,2},x_{I,3})\sigma^z_{I,2,0} 
  &=& H_I^0(-x_{I,2},-x_{I,3}), \\
  \sigma^z_{I,3,0}H_I^0(x_{I,4},x_{I,5})\sigma^z_{I,3,0} 
  &=& H_I^0(-x_{I,4},-x_{I,5}), \\
  \sigma^z_{I,4,0}H_I^0(x_{I,6},x_{I,7})\sigma^z_{I,4,0} 
  &=& H_I^0(-x_{I,6},-x_{I,7}).
\end{eqnarray}
\end{subequations}
Eqs.~(\ref{cornerrules}) correspond to the four corners of the cell, while
Eqs.~(\ref{midptrules}) correspond to the midpoints of each side. The former 
are trivially satisfied if we express the wavefunctions in terms of the $A_k$. 
However, the latter are nontrivial and require that \footnote{
Take, for instance, $\sigma^z_{I,1,0} H_I^0(x_{I,8},x_{I,1})\sigma^z_{I,1,0} 
= H_I^0(-x_{I,8},-x_{I,1})$ and apply it to the state 
$\sigma^z_{I,1,0}\protect\ket{\psi(x_{I,8},x_{I,1})}$, where 
$\protect\ket{\psi(x_{I,8},x_{I,1})}$ is an eigenstate of 
$H_I^0(x_{I,8},x_{I,1})$ with eigenvalue $\epsilon^\text{c}(x_{I,1},x_{I,8})$.
Since this becomes $H_I^0(-x_{I,8},-x_{I,1})(\sigma^z_{I,1,0}
\protect\ket{\psi(x_{I,8},x_{I,1})}) = \epsilon^\text{c}(x_{I,1},x_{I,8})
\sigma^z_{I,1,0}\protect\ket{\psi(x_{I,8},x_{I,1})}$, but 
$\epsilon^\text{c}(x_{I,1},x_{I,8}) = \epsilon^\text{c}(-x_{I,1},-x_{I,8})$,
we must have $\sigma^z_{I,1,0}\protect\ket{\psi(x_{I,8},x_{I,1})} = 
\protect\ket{\psi(-x_{I,8},-x_{I,1})}$.}
\begin{subequations}
\label{midptrule}
\begin{eqnarray}
  \sigma^z_{I,1,0}\ket{i_I(A_4,A_1)}_I &=& \ket{i_I(-A_4,-A_1)}_I, \\ 
  \sigma^z_{I,2,0}\ket{i_I(A_1,A_2)}_I &=& \ket{i_I(-A_1,-A_2)}_I, \\ 
  \sigma^z_{I,3,0}\ket{i_I(A_2,A_3)}_I &=& \ket{i_I(-A_2,-A_3)}_I, \\ 
  \sigma^z_{I,4,0}\ket{i_I(A_3,A_4)}_I &=& \ket{i_I(-A_3,-A_4)}_I.
\end{eqnarray}
\end{subequations}
Note that, in our chosen basis,
\begin{subequations}
\label{internalZ}
\begin{eqnarray}
  \sigma^z_{I,1,0} &=& \mathrm{diag}(+,+,-,-,+,-,+,-), \\
  \sigma^z_{I,2,0} &=& \mathrm{diag}(+,-,+,-,+,-,-,+), \\
  \sigma^z_{I,3,0} &=& \mathrm{diag}(+,+,-,-,-,+,-,+), \\
  \sigma^z_{I,4,0} &=& \mathrm{diag}(+,-,+,-,-,+,+,-). 
\end{eqnarray}
\end{subequations}

\subsubsection{$\Phi_I = +$ sector}

Cell energies arranged in increasing order are
\begin{subequations}
\begin{eqnarray}
  \epsilon_0^\text{c}(+) &=& -2^{3/2}[h^2+J^2+(h^4+J^4)^{1/2}]^{1/2}, 
  \quad\\
  \epsilon_1^\text{c}(+) &=& -2^{3/2}[h^2+J^2-(h^4+J^4)^{1/2}]^{1/2}, \\
  \epsilon_2^\text{c}(+) &=& 0, \\
  \epsilon_3^\text{c}(+) &=& 0, \\
  \epsilon_4^\text{c}(+) &=& 0, \\
  \epsilon_5^\text{c}(+) &=& 0, \\
  \epsilon_6^\text{c}(+) &=& 2^{3/2}[h^2+J^2-(h^4+J^4)^{1/2}]^{1/2}, \\
  \epsilon_7^\text{c}(+) &=& 2^{3/2}[h^2+J^2+(h^4+J^4)^{1/2}]^{1/2}.
  \end{eqnarray}
\end{subequations}
There is no level crossing. For $i_I = 0,1,6,7$, let 
$E = \epsilon_{i_I}^\text{c}(+)$ for brevity. The unnormalized wavefunction is
\begin{subequations}
\label{pluswavefn0}
\begin{equation}
  \ket{0,1,6,7}_I = 
  \begin{pmatrix}
  4J(E+4h)^{-1} \\
  \Gamma(E-4h)A_2A_3 \\
  \Gamma(E-4h)A_3A_4 \\
  \Gamma E A_1A_3 \\
  -A_3 \\
  -A_1 \\
  -A_2 \\
  -A_4
  \end{pmatrix},
\end{equation}
where
\begin{equation}
  \Gamma = \frac{E^2+4hE-4J^2}{J(E+4h)(3E-8h)}.
\end{equation}
For $i_I = 2,3,4,5$, the normalized wavefunctions are 
\begin{eqnarray}
  & & \ket{2}_I =
  \frac{1}{\sqrt{2}}
  \begin{pmatrix}
  0 \\
  A_1 A_4 \\
  -A_3 A_4 \\
  0 \\
  0 \\
  0 \\
  0 \\
  0
  \end{pmatrix}, \quad
  \ket{3}_I =
  \frac{1}{\sqrt{2}}
  \begin{pmatrix}
  0 \\
  0 \\
  0 \\
  0 \\
  A_3 \\
  0 \\
  0 \\
  -A_4
  \end{pmatrix}, \nonumber \\
  & & \ket{4}_I =
  \frac{1}{\sqrt{2}}
  \begin{pmatrix}
  0 \\
  0 \\
  0 \\
  0 \\
  0 \\
  A_1 \\
  -A_2 \\
  0
  \end{pmatrix}, \quad
  \ket{5}_I =
  \frac{1}{2}
  \begin{pmatrix}
  0 \\
  0 \\
  0 \\
  0 \\
  A_3 \\
  -A_1 \\
  -A_2 \\
  A_4
\end{pmatrix}.
\end{eqnarray}
\end{subequations}
It is important to note that $\{\ket{2}_I,\ket{3}_I,\ket{4}_I,\ket{5}_I\}$
form an orthonormal basis in the zero-energy subspace. Obtaining these 
particular wavefunctions required judicious use of identities like 
$A_k = 1/A_k$, and $A_1 = A_2A_3A_4$, $A_1A_2 = A_3A_4$, etc. which follows 
from the fact that $A_1A_2A_3A_4 = 1$.

\subsubsection{$\Phi_I = -$ sector}

Cell energies arranged in increasing order are, for $h > 0$,
\begin{subequations}
\begin{eqnarray}
  \epsilon_0^\text{c}(-) &=& -2(h^2+J^2)^{1/2}-2h \\
  \epsilon_1^\text{c}(-) &=& -2J \\
  \epsilon_2^\text{c}(-) &=& -2J \\
  \epsilon_3^\text{c}(-) &=& -2(h^2+J^2)^{1/2}+2h \\
  \epsilon_4^\text{c}(-) &=& 2(h^2+J^2)^{1/2}-2h \\
  \epsilon_5^\text{c}(-) &=& 2J \\
  \epsilon_6^\text{c}(-) &=& 2J \\
  \epsilon_7^\text{c}(-) &=& 2(h^2+J^2)^{1/2}+2h. 
\end{eqnarray}
\end{subequations}
For $i_I = 0,4,3,7$, let $E = \epsilon^\text{c}_{i_I}(-)$ for brevity. The 
normalized wavefunctions are
\begin{eqnarray}
  & & \ket{0,4}_I = 
  \frac{1}{\sqrt{4+E^2/J^2}}
  \begin{pmatrix}
  -EJ^{-1} \\ 0 \\ 0 \\ 0 \\ A_3 \\ A_1 \\ A_2 \\ A_4
  \end{pmatrix}, \label{minuswavefn} \\
  & & \ket{3,7}_I =
  \frac{1}{\sqrt{4+E^2/J^2}}
  \begin{pmatrix}
  0 \\ 0 \\ 0 \\ -E J^{-1}A_1A_3 \\ A_3 \\ A_1 \\ -A_2 \\ -A_4
  \end{pmatrix}, \nonumber \\
  & & \ket{1}_I =
  \frac{1}{\sqrt{8}}
  \begin{pmatrix}
  0 \\ 2A_2A_3 \\ 0 \\ 0 \\ A_3 \\ -A_1 \\ A_2 \\ -A_4
  \end{pmatrix}, \quad
  \ket{5}_I =
  \frac{1}{\sqrt{8}}
  \begin{pmatrix}
  0 \\ -2A_2A_3 \\ 0 \\ 0 \\ A_3 \\ -A_1 \\ A_2 \\ -A_4
  \end{pmatrix}, \nonumber \\
  & & \ket{2}_I =
  \frac{1}{\sqrt{8}}
  \begin{pmatrix}
  0 \\ 0 \\ 2A_1A_2 \\ 0 \\ -A_3 \\ A_1 \\ A_2 \\ -A_4
  \end{pmatrix}, \quad
  \ket{6}_I = 
  \frac{1}{\sqrt{8}}
  \begin{pmatrix}
  0 \\ 0 \\ -2A_1A_2 \\ 0 \\ -A_3 \\ A_1 \\ A_2 \\ -A_4
  \end{pmatrix}. \nonumber
\end{eqnarray}

\subsection{Lattice eigenstates as products of cell and external link
eigenstates}

For any given configuration on the external links, the lowest energy eigenstate
of $H_\sigma^0$ is obtained from a product over all cells with $i_I = 0$,
\begin{equation}
  \ket{i} = \prod_{I,j} \ket{0(x_{I,1},\dotsc,x_{I,8})}_I\ket{x_{I,j}}_{I,0,j},
\end{equation}
where it is understood that each external link contributes just once. The 
$H_\sigma^0$-eigenvalue is
\begin{equation}
  \epsilon_i = \sum_I\epsilon_0^\text{c}(\Phi_I).
\end{equation}
When summing over all possible $\ket{i}$ we shall use the shorthand
\begin{equation}
  \sum_i = \prod_{I,j} \sum_{x_{I,j} = \pm 1}.
\end{equation}

Corresponding to each state $\ket{i}$ in the Hilbert space of the original 
lattice Hamiltonian is a state $\ket{\mu_i}$ belonging to the smaller Hilbert
space of the renormalized Hamiltonian. Quite simply, it is everything in
$\ket{i}$ but the wavefunction of the internal links.
\begin{equation}
  \ket{\mu_i} = \prod_{I,j}\ket{x_{I,j}}_{I,0,j}.
\end{equation}
It is in this sense that the internal links have been ``decimated.'' On the
thinner lattice of external links $\{\ket{x_{I,j}}_{I,0,j}\}$ we define new 
Pauli operators
\begin{subequations}
\begin{eqnarray}
  \Id_{I,j} &=& (\ket{+}\bra{+} + \ket{-}\bra{-})_{I,0,j}, \\
  \mu^x_{I,j} &=& (\ket{+}\bra{+} - \ket{-}\bra{-})_{I,0,j}, \\
  \mu^z_{I,j} &=& (\ket{+}\bra{-} + \ket{-}\bra{+})_{I,0,j}.
\end{eqnarray}
\end{subequations}

Since $\{\ket{i}\}$ is merely a small subset of the energy basis of 
$H_\sigma^0$, the remaining higher-energy lattice eigenstates are constructed 
from cells with any value of $i_I$,
\begin{equation}
  \ket{\alpha} = \prod_{I,j}\ket{i_I(x_{I,1},\dotsc,x_{I,8})}_I
  \ket{x_{I,j}}_{I,0,j}, \quad i_I = 0,\dotsc,7,
\end{equation}
with the caveat that \emph{at least one} $ i_I > 0$. Its 
$H_\sigma^0$-eigenvalue is
\begin{equation}
  \epsilon_\alpha = \sum_I \epsilon_{i_I}^\text{c}(\Phi_I).
\end{equation}
When summing over all possible $\ket{\alpha}$ we shall use the shorthand
\begin{equation}
  \sum_\alpha = \prod_{I,j} \sum_{i_I = 0}^7\biggr|_\text{some $i_I \neq 0$}
  \sum_{x_{I,j} = \pm 1}.
\end{equation}

\subsection{Hirsch--Mazenko perturbation expansion}

To second order in the intercell coupling,
\begin{subequations}
\begin{eqnarray}
  H^\text{ren}_\mu &=& H_\mu^{(0)} + H_\mu^{(1)} + H_\mu^{(2)} + \dotsb, \\
  H^{(0)}_\mu &=& \sum_i\epsilon_i\ket{\mu_i}\bra{\mu_i}, \label{H0} \\
  H^{(1)}_\mu &=& \sum_{i,i'}\ev{i'|V_\sigma|i}\ket{\mu_i}\bra{\mu_{i'}}, 
  \label{H1} \\
  H^{(2)}_\mu &=& \half\sum_{i,i'}\sum_\alpha
  \ev{i'|V_\sigma|\alpha}\ev{\alpha|V_\sigma|i} \nonumber \\
  & & \times \Bigl(\frac{1}{\epsilon_i-\epsilon_\alpha}+
  \frac{1}{\epsilon_{i'}-\epsilon_\alpha}\Bigr)\ket{\mu_i}\bra{\mu_{i'}}. 
  \label{H2}
\end{eqnarray}
\end{subequations}
Refer to Ref.~\onlinecite{HM} for a derivation of these expressions. They have
been written in a simplified form following Eq.~(3) of Ref.~\onlinecite{Hirsch}.

When there is no chance of confusion, we will abbreviate the state 
$\ket{x_{I,j}}_{I,0,j}$ as $\ket{x_{I,j}}$.

\subsubsection{Computation of $H_\mu^{(0)}$}

Consider Eq.~(\ref{H0}),
\begin{subequations}
\begin{equation}
  H^{(0)}_\mu = 
  \Bigl(\prod_{I,j}\sum_{x_{I,j}}\Bigr)
  \Bigl(\sum_I\epsilon_0^\text{c}(\Phi_I)\Bigr)
  \prod_{I,j}\ket{x_{I,j}}\bra{x_{I,j}}. 
\end{equation}
Bring the sum over cells $I$ out so that only the links belonging to a given
$\Phi_I$ will be non-identity operators. Call the eight links of cell $I$,
$b_1, \dotsc, b_8$. Then
\begin{eqnarray}
  H^{(0)}_\mu &=& 
  \sum_I \sum_{x_{b_1}}\dotsb\sum_{x_{b_8}}
  \epsilon_0^\text{c}(x_{b_1}\dotsb x_{b_8}) \nonumber \\
  & & \times \ket{x_{b_1}}\bra{x_{b_1}}\dotsb\ket{x_{b_8}}\bra{x_{b_8}} \\
  &=& \sum_I \Big(\frac{\epsilon_0^\text{c}(+)+\epsilon_0^\text{c}(-)}{2}
  \Id_{b_1}\dotsb\Id_{b_8} \nonumber \\
  & & - \frac{\epsilon_0^\text{c}(-)-\epsilon_0^\text{c}(+)}{2}
  \mu^x_{b_1}\dotsb\mu^x_{b_8}\Bigr). \label{Jprime}
\end{eqnarray}
\end{subequations}

\subsubsection{Computation of $H_\mu^{(1)}$}

Consider Eq.~(\ref{H1}),
\begin{subequations}
\begin{widetext}
\begin{equation}
  H_\mu^{(1)} = \Bigl(\prod_{I,j}\sum_{x_{I,j}}\Bigr)
  \Bigl(\prod_{I,j}\sum_{x'_{I,j}}\Bigr)
  \prod_{I,j}\bra{0(\{x'_{I,j}\})}_I\bra{x'_{I,j}}
  \cdot -h\sum_{I,j}\sigma^z_{I,0,j}
  \cdot\prod_{I,j}\ket{0(\{x_{I,j}\})}_I \ket{x_{I,j}}
  \cdot\prod_{I,j}\ket{x_{I,j}}\bra{x'_{I,j}}.
\end{equation}
\end{widetext}
By an abuse of notation, each instance of ``$\prod_{I,j}$'' serves to remind us
that there are as many copies of the expression immediately to the right of 
this symbol but left of ``$\cdot$'' or ``$)$'' as external links in the lattice.
Pull out $\sum_{I,j}$. For a single external link at $(I,0,j)$, we get the
constraint $\ev{x'_{I,j}|\sigma^z_{I,0,j}|x_{I,j}} = 
\delta_{x'_{I,j},-x_{I,j}}$. At all other external links the eigenvalues
$x'_{I,j}$ and $x_{I,j}$ are equal. Therefore, all cells $\Itilde$ not
containing link $(I,0,j)$ on their border yield $\ev{0|0}_\Itilde = 1$. 
Only cells $I$ and $\Inn$ share this link. So
\begin{eqnarray}
  H_\mu^{(1)} &=& -h\sum_{I,j} 
  \sum_{x_{I,j}}\sum_{x_{b_1}}\dotsb\sum_{x_{b_{13}}} \nonumber \\
  & & \times 
  \ev{0(-x_{I,j})|0(x_{I,j})}_I \ev{0(-x_{\Inn,[j]})|0(x_{\Inn,[j]})}_\Inn 
  \nonumber \\
  & & \times 
  \ket{x_{I,j}}\bra{-x_{I,j}}\prod_{b=1}^{13}\ket{x_b}\bra{x_b},
\end{eqnarray}
where $b_1,\dotsc,b_{13}$ denote all links in cells $I$ and $\Inn$ besides
$(I,0,j) = (\Inn,0,[j])$. Consider the matrix element 
$\ev{0(-x_{I,j})|0(x_{I,j})}_I$. Besides the choice of $j$ and the value of 
$x_{I,j}$ it could also depend on the seven additional external link 
eigenvalues forming the rest of the boundary of $I$. Let us call them 
$x_{b_1},\dotsc,x_{b_7}$ (we have suppressed writing these as they are not 
negated in the inner product). However, it turns out that this matrix element 
is completely independent of boundary conditions. That is, the matrix element
evaluates to the same quantity for any choice of $j$ and the values 
$x_{I,j},\,x_{b_1},\dotsc,x_{b_7}$. For convenience let us select $j = 1$,
$x_{I,j}=x_{b_1}=\dotsb=x_{b_7}=+$. So the matrix element is an inner product
between the two lowest energy states from the $\Phi_I = +$ and $\Phi_I = -$ 
sectors, respectively. For future convenience define these states to be
(with specific boundary conditions)
\begin{eqnarray}
  \ket{0_+}_I & = & \ket{0(A_1=+,A_2=+,A_3=+,A_4=+)}_I, \label{zeroplus}
  \qquad\qquad \\
  \ket{0_-}_I & = & \ket{0(A_1=+,A_2=-,A_3=+,A_4=+)}_I. \label{zerominus}
\end{eqnarray}
Although it seems peculiar to make $A_2$ negative in Eq.~(\ref{zerominus})
rather than, say, $A_1$, in hindsight this choice allows us to write all matrix
elements using only these two states. We will return to this point later. Thus,
\begin{equation}
  \label{hprime}
  H_\mu^{(1)} = -2h|\ev{0_-|0_+}_I|^2\sum_{I,j}\mu^z_{I,j}.
\end{equation}
\end{subequations}
The factor of 2 arises from the fact that each side of a cell contributes two
links.

\subsubsection{Computation of $H_\mu^{(2)}$}

When fully written out Eq.~(\ref{H2}) is
\begin{widetext}
\begin{eqnarray}
  H_\mu^{(2)} &=& \half
  \Bigl(\prod_{I,j}\sum_{x''_{I,j}}\Bigr)
  \Bigl(\prod_{I,j}\sum_{x'_{I,j}}\Bigr)
  \Bigl(\prod_{I,j}\sum_{i_I=0}^7\Bigr|_\text{some $i_I \neq 0$}
  \sum_{x_{I,j}}\Bigr) \nonumber \\
  & &
  \prod_{I,j}\bra{0(\{x'_{I,j}\})}_I\bra{x'_{I,j}}\cdot 
  -h\sum_{I,j}\sigma^z_{I,0,j}
  \cdot \prod_{I,j}\ket{i_I(\{x_{I,j}\})}_I\ket{x_{I,j}} \nonumber \\
  & & \times
  \prod_{I,j}\bra{i_I(\{x_{I,j}\})}_I\bra{x_{I,j}}\cdot
  -h\sum_{I',j'}\sigma^z_{I',0,j'} \cdot 
  \prod_{I,j}\ket{0(\{x''_{I,j}\})}_I\ket{x''_{I,j}} \nonumber \\
  & &
  \times \Bigl(\frac{1}{\sum_I[\epsilon_0^\text{c}(\Phi''_I) - 
  \epsilon_{i_I}^\text{c}(\Phi_I)]} + 
  \frac{1}{\sum_I[\epsilon_0^\text{c}(\Phi'_I) - 
  \epsilon_{i_I}^\text{c}(\Phi_I)]}
  \Bigr)\cdot
  \prod_{I,j}\ket{x''_{I,j}}\bra{x'_{I,j}},
\end{eqnarray}
\end{widetext}
where 
$\Phi_I = \prod_{j=1}^8 x_{I,j}$, $\Phi'_I = \prod_{j=1}^8 x'_{I,j}$, and
$\Phi''_I = \prod_{j=1}^8 x''_{I,j}$, Pull out $\sum_{I,j}$ and $\sum_{I',j'}$; 
eventually, we will want to keep just one of these sums unevaluated. It is 
possible to completely evaluate $\prod_{I,j}\sum_{x''_{I,j}}$ and 
$\prod_{I,j}\sum_{x'_{I,j}}$ by collapsing Kronecker deltas for external links. 
For links $b \neq (I,0,j)$, $x'_b = x_b$, but for the the special link 
$b = (I,0,j)$, $x'_b = -x_b$. Similarly, for links $b \neq (I',0,j')$, 
$x''_b = x_b$, but for the special link $b = (I',0,j')$, $x''_b = -x_b$. 

Consequently, two kinds of inner product between cell wavefunctions may result.
If a given cell $\Itilde$ does not contain the special link $(I,0,j)$, then 
orthonormality requires that $i_\Itilde = 0$. Likewise, if a given cell $\Ibar$
does not contain the special link $(I',0,j')$, then $i_\Ibar = 0$. This results
in a drastic simplification of the energy denominators. For convenience define
\begin{eqnarray}
  R_{i_I,i_\Inn}(\Phi_I,\Phi_\Inn) 
  & = & \Bigl[
  \epsilon_0^\text{c}(-\Phi_I) + 
  \epsilon_0^\text{c}(-\Phi_\Inn) \nonumber \\
  & & - \epsilon_{i_I}^\text{c}(\Phi_I) -
  \epsilon_{i_\Inn}^\text{c}(\Phi_\Inn)\Bigr]^{-1}. \qquad \label{R}
\end{eqnarray}
However, if a cell does have one of these special links sitting on its 
boundary---there will be two in the case of $(I,0,j)$: $I$ and $\Inn$, and two 
in the case of $(I',0,j')$: $I'$ and $\Ipnn$---then an extra minus sign appears
in one of the eight parameters in one of the two wavefunctions participating in
the inner product. It shall be convenient to define 
\begin{equation}
  \label{zeta}
  \zeta_{i_I}^j(\Phi_I) = 
  \ev{0(-x_{I,j},\{x_{I,k}\}_{k\neq j})|i_I(x_{I,j},\{x_{I,k}\}_{k\neq j})}_I.
\end{equation}
Eq.~(\ref{zeta}) has an important property: although it depends on the choice
of $i_I$ and $j$, it does not depend on the precise choice of the $x_{I,j}$ 
except for the overall sign of $\Phi_I$.

At this step,
\begin{widetext}
\begin{eqnarray}
  H_\mu^{(2)} &=& 
  \frac{h^2}{2}\sum_{I,j}\sum_{I',j'}
  \Bigl(\prod_{I,j}\sum_{i_I}\Bigr|_\text{some $i_I \neq 0$}
  \sum_{x_{I,j}}\Bigr) 
  \prod_{\Itilde \not\ni (I,0,j)} \delta_{0,i_\Itilde}
  \prod_{\Ibar \not\ni (I',0,j')} \delta_{0,i_\Ibar} \nonumber\\
  & & \times \zeta_{i_I}^j(\Phi_I)
  \zeta_{i_\Inn}^{[j]}(\Phi_\Inn)
  \zeta^{j'}_{i_{I'}}(\Phi_{I'})
  \zeta^{[j']}_{i_\Ipnn}(\Phi_\Ipnn) \Bigl(R_{i_I,i_\Inn}(\Phi_I,\Phi_\Inn)
  + R_{i_{I'},i_\Ipnn}(\Phi_{I'},\Phi_\Ipnn)\Bigr) \nonumber\\
  & & \times
  \prod_{b\neq (I,0,j),\,(I',0,j')}
  (\ket{x_b}\bra{x_b})_b 
  \times
  \Biggl\{\begin{array}{ll}
  (\ket{-x_{I,j}}\bra{-x_{I,j}})_{I,0,j} & (I',0,j') = (I,0,j) \\
  (\ket{-x_{I',j'}}\bra{x_{I',j'}})_{I',0,j'}\;
  (\ket{x_{I,j}}\bra{-x_{I,j}})_{I,0,j}  & (I',0,j') \neq (I,0,j) \\
  \end{array}. \quad \label{H2explicit}
\end{eqnarray}
\end{widetext}
If we use up the remaining Kronecker deltas over cells, then the decimation of
internal links will be complete. However, there is not necessarily one
Kronecker delta per cell since the number of such constraints that get enforced 
depends on the relative location of link $(I',0,j')$ to link $(I,0,j)$. 
For imagine that $(I,0,j)$ is fixed. If $(I',0,j')$ is located at...
\begin{enumerate}
\item
Any of the two links on the shared boundary of $I$ and $\Inn$, then $I'$ and 
$\Ipnn$ coincide with $I$ and $\Inn$ precisely. Therefore, Kronecker deltas 
will not exist for these two cells. And $\sum_{i_I}$ and $\sum_{i_\Inn}$ will 
be left undone;
\item
Any of the twelve links on the outer perimeter of the union of $I$ and $\Inn$, 
then either $I'$ or $\Ipnn$ will coincide with one of $I$ and $\Inn$. 
Therefore, a Kronecker delta will not exist for that doubly-covered cell. Say 
this is $I$. Then $\sum_{i_I}$ will be left undone;
\item
Any other link on the lattice, then $I'$ and $\Ipnn$ will not overlap $I$ and 
$\Inn$ at all. Therefore, a Kronecker delta exists for all cells.
\end{enumerate}
In terms of these three general cases, let us write
\begin{equation}
  \label{H2cased}
  H^{(2)}_\mu = \frac{h^2}{2}\sum_{I,j}
  (\text{Case 1}+\text{Case 2}+\text{Case 3}).
\end{equation}

\paragraph{Case 3}

If every cell is constrained to be in its ground state, then there cannot be
an intermediate excited state (i.e., $\sum_{i_I}|_\text{some $i_I \neq 0$}$ 
is null). Thus,
\begin{equation}
  \text{Case 3} = 0.
\end{equation}

\paragraph{Case 1}

\begin{figure}
\includegraphics{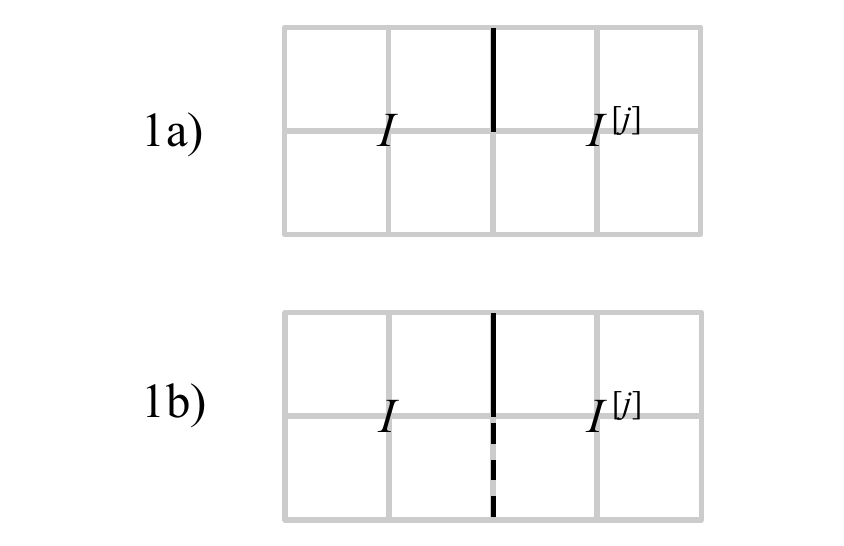}
\caption{\label{fig:case1}
Subcases 1a and 1b. Links $(I,0,j)$ and $(I',0,j')$ are represented by the
bold and dashed links, respectively. In Subcase 1a they are the same link.}
\end{figure}

There are two subcases: (a) $(I',0,j') = (I,0,j)$, and
(b) $(I',0,j') \neq (I,0,j)$, such that
\begin{equation}
  \text{Case 1} = \text{Subcase 1a} + \text{Subcase 1b}.
\end{equation}
See Fig.~\ref{fig:case1}.
For all external links that are not on the two neighboring cells that share 
the links $(I,0,j)$ and $(I',0,j')$, we obtain the identity operator since
$\sum_{x_b}(\ket{x_b}\bra{x_b})_b = \Id_b$. There are fourteen link sums left 
to do.

Consider Subcase 1a. Define
\begin{eqnarray}
  C^{j,i_I,i_\Inn}(\Phi_I,\Phi_\Inn) &=& \zeta_{i_I}^j(\Phi_I)^2
  \zeta_{i_\Inn}^{[j]}(\Phi_\Inn)^2 \nonumber \\
  & & \times 2R_{i_I,i_\Inn}(\Phi_I,\Phi_\Inn). \quad
\end{eqnarray}
This is, essentially, the second line in Eq.~(\ref{H2explicit}).
Summing over the fourteen external links and using the identity 
$\mu^x_b\mu^x_b = \Id_b$ gives
\begin{eqnarray}
  \text{Subcase 1a} 
  &=& \Biggl(\sum_{i_\Inn=1}^7\Bigr|_{i_I=0} + 
  \sum_{i_I=1}^7\Bigr|_{i_\Inn=0} + 
  \sum_{i_I=1}^7\sum_{i_\Inn=1}^7\Biggr) \nonumber \\
  & & \Biggl[
  S_{++++}^{j,i_I,i_\Inn} \prod_{b\in I}\Id_b \cdot \prod_{b\in\Inn}\Id_b
  \nonumber \\
  & & 
  + S_{+--+}^{j,i_I,i_\Inn} \prod_{b\in I}\mu_b^x \cdot 
  \prod_{b\in\Inn} \mu_b^x \nonumber \\
  & & 
  - S_{++--}^{j,i_I,i_\Inn} \prod_{b\in I}\mu_b^x \cdot \prod_{b\in\Inn}\Id_b
  \nonumber \\
  & & 
  - S_{+-+-}^{j,i_I,i_\Inn} \prod_{b\in I}\Id_b \cdot \prod_{b\in\Inn} 
  \mu_b^x\Biggr],
\end{eqnarray}
where
\begin{eqnarray}
  S_{\sigma_1\sigma_2\sigma_3\sigma_4}^{j,i_I,i_\Inn} & = &
  \frac{1}{4}\Bigl[
  \sigma_1 C^{j,i_I,i_\Inn}(+,+) \nonumber \\
  & & + \sigma_2 C^{j,i_I,i_\Inn}(+,-) \nonumber \\
  & & + \sigma_3 C^{j,i_I,i_\Inn}(-,+) \nonumber \\
  & & + \sigma_4 C^{j,i_I,i_\Inn}(-,-)\Bigr]. 
\end{eqnarray}
Further consolidation is achieved by defining
\begin{subequations}
\begin{eqnarray}
  S_1 & = & 
  \sum_{i_\Inn=1}^7 S_{++++}^{j,0,i_\Inn} 
  + \sum_{i_I=1}^7 S_{++++}^{j,i_I,0} \nonumber \\
  & & + \sum_{i_I=1}^7\sum_{i_\Inn=1}^7 S_{++++}^{j,i_I,i_\Inn}, \\
  S_2 & = &
  \sum_{i_\Inn=1}^7 S_{+--+}^{j,0,i_\Inn} 
  + \sum_{i_I=1}^7 S_{+--+}^{j,i_I,0} \nonumber \\
  & & + \sum_{i_I=1}^7\sum_{i_\Inn=1}^7 S_{+--+}^{j,i_I,i_\Inn}, \\
  S_3 & = & 
  \sum_{i_\Inn=1}^7 S_{++--}^{j,0,i_\Inn} 
  + \sum_{i_I=1}^7 S_{++--}^{j,i_I,0} \nonumber \\
  & & + \sum_{i_I=1}^7\sum_{i_\Inn=1}^7 S_{++--}^{j,i_I,i_\Inn} \label{S3a} \\
  & = & \sum_{i_\Inn=1}^7 S_{+-+-}^{j,0,i_\Inn} 
  + \sum_{i_I=1}^7 S_{+-+-}^{j,i_I,0} \nonumber \\
  & & + \sum_{i_I=1}^7\sum_{i_\Inn=1}^7 S_{+-+-}^{j,i_I,i_\Inn}. \label{S3b}
\end{eqnarray}
\end{subequations}
We have checked that expressions (\ref{S3a}) and (\ref{S3b}) are equivalent. 
Furthermore, none of the expressions for $S_1$, $S_2$, and $S_3$ depend on the 
choice of parameter $j$. This is expected since the lattice remains unchanged
by $90^\circ$ rotations, or reflections about a horizontal or vertical line. 
Letting $\Phi_I = \prod_{b\in I}\mu_b^x$, we get
\begin{eqnarray}
  \text{Subcase 1a} & = & S_1 + S_2\Phi_I\Phi_\Inn \nonumber \\
  & & - S_3(\Phi_I + \Phi_\Inn).
\end{eqnarray}

Next consider Subcase 1b. Define
\begin{eqnarray}
  D^{j,i_I,i_\Inn}(\Phi_I,\Phi_\Inn) & = &
  \zeta_{i_I}^j(\Phi_I)
  \zeta_{i_\Inn}^{[j]}(\Phi_\Inn)
  \zeta_{i_I}^{j'}(\Phi_I)
  \zeta_{i_\Inn}^{[j']}(\Phi_\Inn) \nonumber \\
  & & \times 2R_{i_I,i_\Inn}(\Phi_I,\Phi_\Inn).
  \end{eqnarray}
Summing over the fourteen external links and using the identity 
$(\ket{-}\bra{+} - \ket{+}\bra{-})_b = \mathrm{i}\mu^y_b = \mu^z_b\mu^x_b$
gives
\begin{eqnarray}
  \text{Subcase 1b} &=& 
  \Biggl(\sum_{i_\Inn=1}^7\Bigr|_{i_I=0} + 
  \sum_{i_I=1}^7\Bigr|_{i_\Inn=0} + 
  \sum_{i_I=1}^7\sum_{i_\Inn=1}^7\Biggr) \nonumber \\
  & & 
  \mu_{I,j}^z\mu_{I,j'}^z
  \Biggl[
  T_{++++}^{j,i_I,i_\Inn}\prod_{b\in I}\Id_b \cdot \prod_{b\in\Inn}\Id_b 
  \nonumber \\
  & & 
  + T_{+--+}^{j,i_I,i_\Inn}\prod_{b\in I}\mu_b^x \cdot \prod_{b\in\Inn} \mu_b^x
  \nonumber \\
  & & 
  - T_{++--}^{j,i_I,i_\Inn}\prod_{b\in I}\mu_b^x \cdot \prod_{b\in\Inn}\Id_b
  \nonumber \\
  & & 
  - T_{+-+-}^{j,i_I,i_\Inn}\prod_{b\in I}\Id_b \cdot \prod_{b\in\Inn}\mu_b^x
  \Biggr],
\end{eqnarray}
where
\begin{eqnarray}
  T_{\sigma_1\sigma_2\sigma_3\sigma_4}^{j,i_I,i_\Inn} & = &
  \frac{1}{4}\Bigl[
  \sigma_1 D^{j,i_I,i_\Inn}(+,+) \nonumber \\
  & & + \sigma_2 D^{j,i_I,i_\Inn}(+,-) \nonumber \\
  & & + \sigma_3 D^{j,i_I,i_\Inn}(-,+) \nonumber \\
  & & + \sigma_4 D^{j,i_I,i_\Inn}(-,-)\Bigr]. 
\end{eqnarray}
Once again, further consolidation is achieved by defining
\begin{subequations}
\begin{eqnarray}
  T_1 & = & 
  \sum_{i_\Inn=1}^7 T_{++++}^{j,0,i_\Inn} 
  + \sum_{i_I=1}^7 T_{++++}^{j,i_I,0} \nonumber \\
  & & + \sum_{i_I=1}^7\sum_{i_\Inn=1}^7 T_{++++}^{j,i_I,i_\Inn}, \\
  T_2 & = &
  \sum_{i_\Inn=1}^7 T_{+--+}^{j,0,i_\Inn} 
  + \sum_{i_I=1}^7 T_{+--+}^{j,i_I,0} \nonumber \\
  & & + \sum_{i_I=1}^7\sum_{i_\Inn=1}^7 T_{+--+}^{j,i_I,i_\Inn}, \\
  T_3 & = & 
  \sum_{i_\Inn=1}^7 T_{++--}^{j,0,i_\Inn} 
  + \sum_{i_I=1}^7 T_{++--}^{j,i_I,0} \nonumber \\
  & & + \sum_{i_I=1}^7\sum_{i_\Inn=1}^7 T_{++--}^{j,i_I,i_\Inn} \label{T3a} \\
  & = & \sum_{i_\Inn=1}^7 T_{+-+-}^{j,0,i_\Inn} 
  + \sum_{i_I=1}^7 T_{+-+-}^{j,i_I,0} \nonumber \\
  & & + \sum_{i_I=1}^7\sum_{i_\Inn=1}^7 T_{+-+-}^{j,i_I,i_\Inn}. \label{T3b}
\end{eqnarray}
\end{subequations}
We have checked that expressions (\ref{T3a}) and (\ref{T3b}) are equivalent, and
that none of the expressions for $T_1$, $T_2$, and $T_3$ depend on the choice 
of parameter $j$. Thus,
\begin{eqnarray}
  \text{Subcase 1b} & = & \mu^z_{I,j}\mu^z_{I,j'}[T_1 + T_2\Phi_I\Phi_\Inn 
  \nonumber \\
  & & - T_3(\Phi_I + \Phi_\Inn)].
\end{eqnarray}

\paragraph{Case 2}

\begin{figure*}
\includegraphics{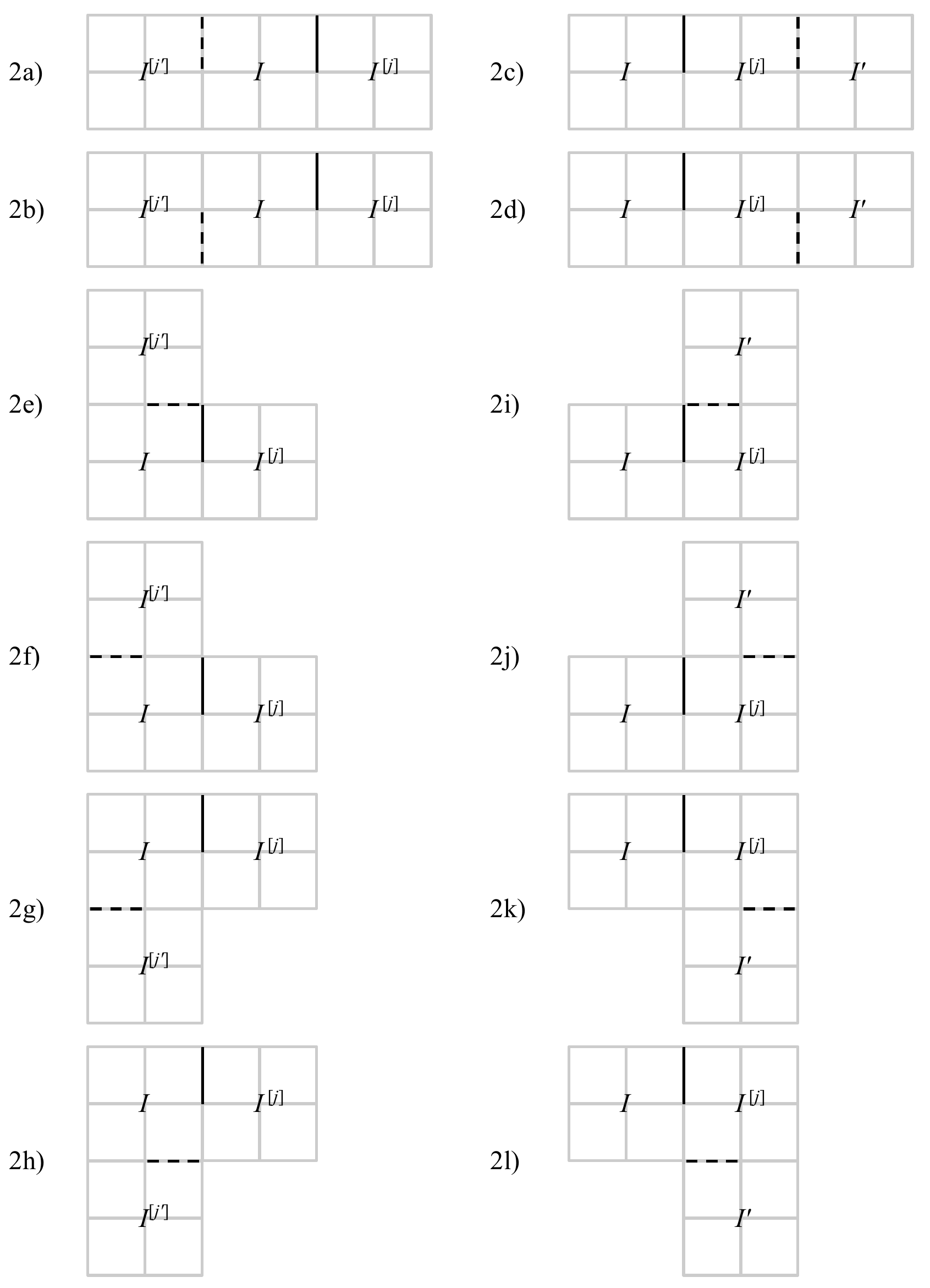}
\caption{\label{fig:case2}
Subcases 2a through 2l. Links $(I,0,j)$ and $(I',0,j')$ are represented by the
bold and dashed links, respectively.}
\end{figure*}

There are twelve subcases such that
\begin{equation}
  \text{Case 2} = \text{Subcase 2a} + \dotsb + \text{Subcase 2l}.
\end{equation}
Each subcase falls naturally into one of two groups based on the location of 
$(I',0,j')$ on the perimeter of the rectangular region formed by $I$ and $\Inn$:
(a)--(d) have $(I',0,j')$ as one of the four links on the short sides 
of the rectangle; (e)--(l) have $(I',0,j')$ as one of the eight links on the 
long sides of the rectangle. See Fig.~\ref{fig:case2}. 

Consider Subcases 2a, b, e, f, g, h. It is possible to regard the link 
$(I',0,j')$ as $(I,0,j')$ if we identify cell $I'$ with $I$. Define
\begin{eqnarray}
  & & E^{j,j',i_I}(\Phi_I,\Phi_\Inn,\Phi_\Innp) \nonumber \\
  & = & \zeta^j_{i_I}(\Phi_I) \zeta^{[j]}_0(\Phi_\Inn)
  \zeta^{j'}_{i_I}(\Phi_I) \zeta^{[j']}_0(\Phi_\Innp) \nonumber \\
  & & \times
  \Bigl[R_{i_I,0}(\Phi_I,\Phi_\Inn) + R_{i_I,0}(\Phi_I,\Phi_\Innp)\Bigr]. \quad
\end{eqnarray}
Summing over the twenty external links gives
\begin{widetext}
\begin{eqnarray}
  & & \text{Subcase 2a, b, e, f, g, h} \nonumber \\ 
  & = & 
  \sum_{i_I = 1}^7 \mu_{I,j}^z\mu_{I,j'}^z \Biggl[
  U^{j,j',i_I}_{++++|++++}
  \prod_{b\in\Innp}\Id_b \cdot \prod_{b\in I}\Id_b \cdot \prod_{b\in\Inn}\Id_b
  +U^{j,j',i_I}_{+-+-|+-+-}
  \prod_{b\in\Innp}\mu_b^x \cdot \prod_{b\in I}\Id_b \cdot \prod_{b\in\Inn}\Id_b
  \nonumber \\
  & & 
  -U^{j,j',i_I}_{++++|----}
  \prod_{b\in\Innp}\Id_b \cdot \prod_{b\in I}\mu^x_b \cdot \prod_{b\in\Inn}\Id_b
  -U^{j,j',i_I}_{+-+-|-+-+}
  \prod_{b\in\Innp}\mu^x_b \cdot \prod_{b\in I}\mu^x_b \cdot \prod_{b\in\Inn}
  \Id_b \nonumber \\
  & &
  -U^{j,j',i_I}_{++--|++--}
  \prod_{b\in\Innp}\Id_b \cdot \prod_{b\in I}\Id_b \cdot \prod_{b\in\Inn}\mu^x_b
  -U^{j,j',i_I}_{+--+|+--+}
  \prod_{b\in\Innp}\mu^x_b \cdot \prod_{b\in I}\Id_b \cdot \prod_{b\in\Inn}
  \mu^x_b \nonumber \\
  & & 
  +U^{j,j',i_I}_{++--|--++}
  \prod_{b\in\Innp}\Id_b \cdot \prod_{b\in I}\mu^x_b \cdot \prod_{b\in\Inn}
  \mu^x_b
  +U^{j,j',i_I}_{+--+|-++-}
  \prod_{b\in\Innp}\mu^x_b \cdot \prod_{b\in I}\mu^x_b \cdot
  \prod_{b\in\Inn}\mu^x_b\Biggr],
\end{eqnarray}
\end{widetext}
where
\begin{eqnarray}
  U^{j,j',i_I}_{\sigma_1\sigma_2\sigma_3\sigma_4|\sigma_5\sigma_6\sigma_7
  \sigma_8} & = & \frac{1}{8}\Bigl[
  \sigma_1 E^{j,j',i_I}(+,+,+) \nonumber \\
  & & +\sigma_2 E^{j,j',i_I}(+,+,-) \nonumber \\
  & & +\sigma_3 E^{j,j',i_I}(+,-,+) \nonumber \\
  & & +\sigma_4 E^{j,j',i_I}(+,-,-) \nonumber \\
  & & +\sigma_5 E^{j,j',i_I}(-,+,+) \nonumber \\
  & & +\sigma_6 E^{j,j',i_I}(-,+,-) \nonumber \\
  & & +\sigma_7 E^{j,j',i_I}(-,-,+) \nonumber \\
  & & +\sigma_8 E^{j,j',i_I}(-,-,-) \Bigr]. \quad
\end{eqnarray}
Define 
\begin{subequations}
\begin{eqnarray}
  U_1^{j,j'} & = & \sum_{i_I=1}^7 U^{j,j',i_I}_{++++|++++} \\
  U_2^{j,j'} & = & \sum_{i_I=1}^7 U^{j,j',i_I}_{+-+-|+-+-} \label{U2a} \\
  & = & \sum_{i_I=1}^7 U^{j,j',i_I}_{++--|++--} \label{U2b} \\
  U_3^{j,j'} & = & \sum_{i_I=1}^7 U^{j,j',i_I}_{++++|----} \\
  U_4^{j,j'} & = & \sum_{i_I=1}^7 U^{j,j',i_I}_{+-+-|-+-+} \label{U4a} \\
  & = & \sum_{i_I=1}^7 U^{j,j',i_I}_{++--|--++}. \label{U4b}
\end{eqnarray}
\end{subequations}
We have checked that expressions (\ref{U2a}) and (\ref{U2b}), and expressions
(\ref{U4a}) and (\ref{U4b}) are equivalent. Also,
\begin{subequations}
\begin{eqnarray}
  \sum_{i_I=1}^7 U^{j,j',i_I}_{+--+|+--+} & = & 0 \\
  \sum_{i_I=1}^7 U^{j,j',i_I}_{+--+|-++-} & = & 0. 
\end{eqnarray}
\end{subequations}
Thus,
\begin{eqnarray}
  & & \text{Subcase 2a, b, e, f, g, h} \nonumber \\ 
  & = & \mu^z_{I,j}\mu^z_{I,j'}[
  U_1^{j,j'} \nonumber \\
  & & + U_2^{j,j'}\Phi_\Innp
  - U_2^{j,j'}\Phi_\Inn
  - U_3^{j,j'}\Phi_I \nonumber \\
  & & - U_4^{j,j'}\Phi_\Innp\Phi_I 
  + U_4^{j,j'}\Phi_I\Phi_\Inn].
  \label{subcase2abefgh}
\end{eqnarray}

Notice that the mapping
\begin{eqnarray}
  j & \to & [j] \nonumber \\
  j' & \to & [j'] \nonumber \\
  I & \to & \Inn \nonumber \\
  \Inn & \to & I \nonumber \\
  \Innp & \to & I' \label{rules}
\end{eqnarray}
converts the diagrams for Subcases 2a, b, e, f, g, and h into the diagrams for
Subcases 2c, d, i, j, k, and l, respectively, up to a reflection in the plane.
However, this reflection does not affect the mathematical expression since it
does not alter the relative locations of links. Applying rules~(\ref{rules}) to
Eq.~(\ref{subcase2abefgh}) and remembering that now we cannot identify cells 
$I'$ and $I$, yields
\begin{eqnarray}
  & & \text{Subcase 2c, d, i, j, k, l} \nonumber \\ 
  & = & \mu^z_{I,j}\mu^z_{I',j'}[
  U_1^{[j],[j']} \nonumber \\
  & & + U_2^{[j],[j']}\Phi_{I'}
  - U_2^{[j],[j']}\Phi_I
  - U_3^{[j],[j']}\Phi_\Inn \nonumber \\ 
  & & - U_4^{[j],[j']}\Phi_{I'}\Phi_\Inn 
  + U_4^{[j],[j']}\Phi_\Inn\Phi_I].
\end{eqnarray}

\subsection{Renormalized spin-$\half$ operators}

Since each link $B$ of the new lattice corresponds to two links of a cell, say
$b$ and $b'$, a prescription is needed to define new link operators 
$\{X_B,\,Z_B\}$ from the old ones $\{\mu^x_b,\,\mu^z_b\}$ and 
$\{\mu^x_{b'},\,\mu^z_{b'}\}$. Although it appears that external links $b$ and 
$b'$ have two qubits worth of freedom, Fradkin and Raby showed that gauge 
invariance restricts this freedom to just one. Here we translate their argument
into a format that comports with the perturbative framework of Hirsch and 
Mazenko.

Consider the site at the midpoint of the boundary between cells $I$ and 
$I+\xhat$ in Fig.~\ref{fig:cellnotate}. The generator of gauge transformations 
at this site is $G_\sigma = \sigma^z_{I,0,1}\sigma^z_{I,0,8}
\sigma^z_{I,1,0}\sigma^z_{I+\xhat,3,0}$. Physical states in Hilbert space must 
satisfy $G_\sigma = +1$. Since $[G_\sigma, H_\sigma^0] = 0$, the 
bi-vector-valued quantity $T_0 = \sum_i\ket{i}\otimes\ket{\mu_i}$ must satisfy
\begin{equation}
  T_0 = G_\sigma T_0.
\end{equation}
Whence does $T_0$ come? Briefly, according to Ref.~\onlinecite{HM}, $T_0$ is 
the lowest-order-in-$g$ approximation to $T[\mu|\sigma]$, which is a 
vector-to-vector projection operator that allows the renormalized Hamiltonian
to be computed by a trace. \footnote{
In Eq.~(2.3) of Ref.~\onlinecite{HM}, the projection is written as 
$H^\text{ren}_\mu = \Tr_\sigma(H_\sigma T[\mu|\sigma]T^\dag[\mu|\sigma])$.}
In Eq.~(2.5) $T_0$ is seen to satisfy a normalization constraint, 
$\Tr_\sigma(T_0 T_0^\dag) = \Id_\mu$. This is equivalent to 
\begin{equation}
  \Id_\mu = \Tr_\sigma(G_\sigma T_0 T_0^\dag) = 
  \sum_{i,i'}\ev{i'|G_\sigma|i}\ket{\mu_i}\bra{\mu_i},
\end{equation}
with the second equality following from $\ev{i|\alpha} = 0$. Simply put, the 
identity operator in the reduced Hilbert space can be computed by restricting 
$G_\sigma$ to the subspace $\{\ket{i}\}$ of lattice eigenstates formed by cell 
ground states; the orthogonal subspace $\{\ket{\alpha}\}$ is completely 
overlooked by the projector $T_0$. Using Eq.~(\ref{midptrule}),
\begin{eqnarray}
  G_\sigma\ket{i} &=& 
  \ket{0(-x_{I,1},-x_{I,8})}_I
  \ket{0(-x_{I+\xhat,4},-x_{I+\xhat,5})}_{I+\xhat} \nonumber \\ 
  & &
  \ket{-x_{I,1}}_{I,0,1}\ket{-x_{I,8}}_{I,0,8}\dotsb, \label{Gi}
\end{eqnarray}
where ellipses represent cell and external link states that are not acted upon 
by $G_\sigma$. Note that $x_{I+\xhat,4} = x_{I,1}$ and 
$x_{I+\xhat,5} = x_{I,8}$. When $\bra{i'}$ is applied to Eq.~(\ref{Gi}),
the resulting matrix element is nonzero only if $x'_{I,0,1} = -x_{I,0,1}$ and
$x'_{I,0,8} = -x_{I,0,8}$, with all other $x'_b = x_b$. If this holds, then
the matrix element is unity. This constraint collapses the sum over $i'$. 
Performing the remaining sum over $i$ then leads to 
\begin{equation}
  \sum_{x_{I,1}=\pm}\ket{x_{I,1}}\bra{-x_{I,1}}
  \sum_{x_{I,8}=\pm}\ket{x_{I,8}}\bra{-x_{I,8}}
  = \mu^z_{I,1}\mu^z_{I,8},
\end{equation}
with implicit identities on all other external links. Thus, we have proven that 
\begin{equation}
  \mu^z_{I,1}\mu^z_{I,8} = \Id_\mu.
\end{equation}
Therefore, $\mu^z_{I,1} = \mu^z_{I,8}$ on the reduced Hilbert space of the
thinned lattice.

Define renormalized link operators
\begin{equation}\label{renormlinkops}
  X_B = \mu^x_b \mu^x_{b'}, \qquad Z_B = (\mu^z_b + \mu^z_{b'})/2,
\end{equation}
where $b$ and $b'$ are the two contiguous links from the same edge $B$ of a 
cell. See Fig.~\ref{fig:H2simple}. It is easily checked that they reproduce the 
Pauli algebra $X_B Z_B = -Z_B X_B$, $X_B^2 = \Id_B$, and $Z_B^2 = \Id_B$. Note 
that the identity $\mu^z_b \mu^z_{b'} = \Id_B$ implies that 
$Z_B = \mu_b^z = \mu_{b'}^z$.

Cells become plaquettes in the thinned lattice. For these we define the magnetic
flux in the obvious way,
\begin{equation}
  \Phi_I = \prod_{B \in \partial I} X_B.
\end{equation}

\subsection{Renormalized Hamiltonian}

In terms of the renormalized spin-$\half$ operators, Eq.~(\ref{Jprime}) reads
\begin{equation}
\label{H0solved}
  H^{(0)}_\mu = \frac{\epsilon_0^\text{c}(+)+\epsilon_0^\text{c}(-)}{2}
  \frac{N_\text{plaq}}{4} 
  - \frac{\epsilon_0^\text{c}(-)-\epsilon_0^\text{c}(+)}{2}\sum_I\Phi_I,
\end{equation}
where $N_\text{plaq}$ is the number of plaquettes in the original lattice. And 
for Eq.~(\ref{hprime}) we make a minor notational change: instead of referring 
to links as ``$I,j$'' we use ``$B$,'' 
\begin{equation}
\label{H1solved}
  H^{(1)}_\mu = -2h|\ev{0_-|0_+}_I|^2\sum_B Z_B.
\end{equation}

\begin{figure}
\includegraphics{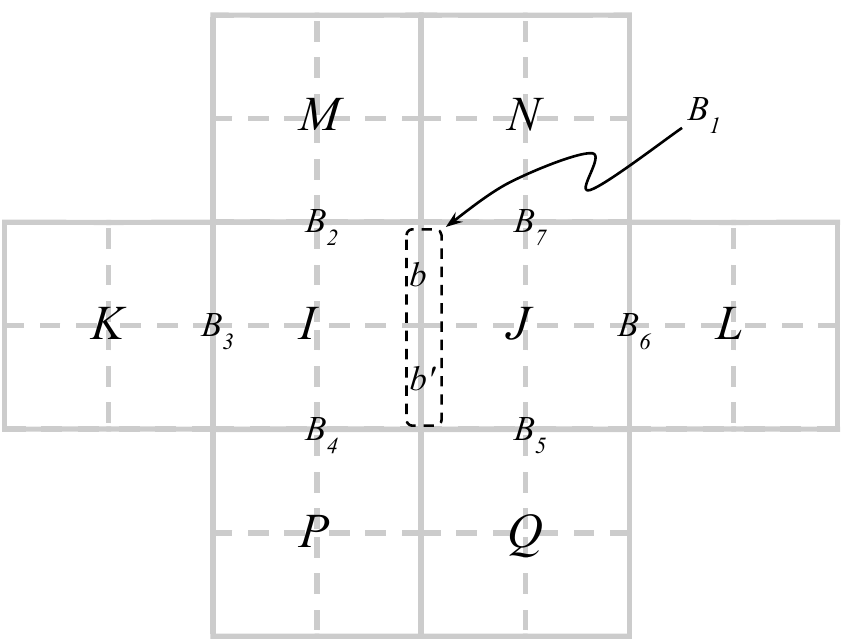}
\caption{\label{fig:H2simple}
Cells involved in effective operators generated at second order in the 
intercell coupling. Cells $I,\,J,\,K,\,L,\,M,\,N,\,P,\,Q$ become the plaquettes
of the thinned lattice once internal links (dashed) are decimated. External 
links (solid) $b$ and $b'$ comprise the link $B_1$ in the thinned lattice.}
\end{figure}

We are not quite ready to write a gauge-invariant expression for $H^{(2)}_\mu$.
First, Eq.~(\ref{H2cased}) needs to be assembled by combining all subcases from 
Cases 1 and 2. Since pairs of contiguous links that form the edge of a cell 
will eventually become a single link $B$ in the thinned lattice, it is 
convenient to reduce the scope of the sum as 
\begin{equation}
  H^{(2)}_\mu = \frac{h^2}{2}\sum_B 
  (\text{Case 1} + \text{Case 2})_{j = 1} 
  + (\text{Case 1} + \text{Case 2})_{j = 8}. 
\end{equation}
Note that we could have also picked the pair $j = 2,3$, or $4,5$, or $6,7$.
However, this choice is immaterial due to the $90^\circ$ rotation symmetry of
the lattice. Also, the choice $j = 1,8$ matches the links labeled $b,b'$ in 
Fig.~\ref{fig:H2simple} whose simplified notation we henceforth employ. 
Accordingly, for Case 1 we have
$\text{Subcase 1a} = S_1 + S_2\Phi_I\Phi_J - S_3(\Phi_I + \Phi_J)$ and 
$\text{Subcase 1b} = T_1 + T_2\Phi_I\Phi_J - T_2(\Phi_I + \Phi_J)$ because
$\mu^z_b\mu^z_{b'} = \Id_B$. For Case 2, we must be careful to properly overlay
Fig.~\ref{fig:H2simple} onto each diagram shown in Fig.~\ref{fig:case2} so that
the cells labeled ``$I$'' coincide. For instance, consider Subcase 2c with 
$j = 1$. Then $\Inn = J$, $I' = L$, and $j' = 4$. Therefore,
\begin{eqnarray}
  (\text{Subcase 2c})_{j=1} & = & Z_{B_1}Z_{B_6}[U_1^{[1],[4]} \nonumber \\
  & & + U_2^{[1],[4]}\Phi_L
  - U_2^{[1],[4]}\Phi_I
  - U_3^{[1],[4]}\Phi_J \nonumber \\ 
  & & - U_4^{[1],[4]}\Phi_L\Phi_J 
  + U_4^{[1],[4]}\Phi_J\Phi_I]. \quad\qquad
\end{eqnarray}
By using relations (\ref{bracketmap}) we can set $[1] = 4$ and $[4] = 1$.
For another example, consider Subcase 2f with $j = 8$. The diagram for this 
may be obtained by reflecting diagram 2f in Fig.~\ref{fig:case2} about a 
horizontal line. Overlaying Fig.~\ref{fig:H2simple} then gives $\Inn = J$, 
$\Innp = P$, and $j' = 6$. Therefore,
\begin{eqnarray}
  (\text{Subcase 2f})_{j=8} & = & Z_{B_1}Z_{B_4}[U_1^{8,6} \nonumber \\
  & & + U_2^{8,6}\Phi_P
  - U_2^{8,6}\Phi_J
  - U_3^{8,6}\Phi_I \nonumber \\ 
  & & - U_4^{8,6}\Phi_P\Phi_I 
  + U_4^{8,6}\Phi_I\Phi_J]. \qquad
\end{eqnarray}

In order to combine all subcases under Case 2 define
\begin{eqnarray}
  V_k 
  & = & U_k^{1,2} + U_k^{1,3} + U_k^{8,3} + U_k^{8,2} \\
  & = & U_k^{1,6} + U_k^{1,7} + U_k^{8,7} + U_k^{8,6} \nonumber \\
  & = & U_k^{4,3} + U_k^{4,2} + U_k^{5,2} + U_k^{5,3} \nonumber \\
  & = & U_k^{4,7} + U_k^{4,6} + U_k^{5,6} + U_k^{5,7}, \nonumber \\
  W_k 
  & = & U_k^{1,4} + U_k^{1,5} + U_k^{8,5} + U_k^{8,4} \\
  & = & U_k^{4,1} + U_k^{4,8} + U_k^{5,8} + U_k^{5,1}, \nonumber
\end{eqnarray}
for $k = 1,\dotsc, 4$. Notice that $V_k$ are coefficients from terms in which 
$Z$'s reside on nearest-neighbor (i.e., diagonally adjacent) links, whereas 
$W_k$ are coefficients from terms in which $Z$'s reside on next-nearest-neighbor
(i.e., directly opposite) links. The contribution from link $B_1$ is
\begin{widetext}
\begin{eqnarray}
  (\text{Case 1}+\text{Case 2})_{B_1} & = & 
  2(S_1+T_1) + 2(S_2+T_2)\Phi_I\Phi_J - 2(S_3+T_3)(\Phi_I+\Phi_J) \nonumber \\
  & & +Z_{B_1}Z_{B_2}[V_1 + V_2\Phi_M - V_2\Phi_J - V_3\Phi_I - V_4\Phi_M\Phi_I
  + V_4\Phi_I\Phi_J] \nonumber \\
  & & +Z_{B_1}Z_{B_4}[V_1 + V_2\Phi_P - V_2\Phi_J - V_3\Phi_I - V_4\Phi_P\Phi_I
  + V_4\Phi_I\Phi_J] \nonumber \\
  & & +Z_{B_1}Z_{B_5}[V_1 + V_2\Phi_Q - V_2\Phi_I - V_3\Phi_J - V_4\Phi_Q\Phi_J
  + V_4\Phi_J\Phi_I] \nonumber \\
  & & +Z_{B_1}Z_{B_7}[V_1 + V_2\Phi_N - V_2\Phi_I - V_3\Phi_J - V_4\Phi_N\Phi_J
  + V_4\Phi_J\Phi_I] \nonumber \\
  & & +Z_{B_1}Z_{B_3}[W_1 + W_2\Phi_K - W_2\Phi_J - W_3\Phi_I - W_4\Phi_K\Phi_I
  + W_4\Phi_I\Phi_J] \nonumber \\
  & & +Z_{B_1}Z_{B_6}[W_1 + W_2\Phi_L - W_2\Phi_I - W_3\Phi_J - W_4\Phi_L\Phi_J
  + W_4\Phi_J\Phi_I].
\end{eqnarray}
\end{widetext}
Observe that not all of these terms are hermitian. For example, 
$(Z_{B_1}Z_{B_3}\Phi_K)^\dag = \Phi_K Z_{B_3} Z_{B_1} = -Z_{B_1}Z_{B_3}\Phi_K$
because $X_{B_3}$ anticommutes with $Z_{B_3}$. Such terms must disappear when 
we add similar contributions from the other links. For instance, if we look at
$(\text{Case 1}+\text{Case 2})_{B_3}$ then we also get a term prefaced with 
$Z_{B_3}Z_{B_1}$. Combining like terms eliminates non-hermitian operators. In
particular, operators with coefficients $V_k$ and $W_k$ for $k = 2, 4$ vanish.

For any pair of nearest-neighbor links $B$ and $B'$ at right angles (denoted
$B \perp B'$), $H^{(2)}_\mu$ contains operators of the form
\begin{equation}
  h^2 Z_B Z_{B'} (V_1 - V_3\Phi_I),
\end{equation}
where plaquette $I$ is bounded by links $B$ and $B'$. We might say that $I$ 
sits in the ``elbow'' of the hook made by $B$ and $B'$.

For any pair of next-nearest-neighbor links $B$ and $B'$ directly opposite from
each other (denoted $B \parallel B'$), $H^{(2)}_\mu$ contains operators of the 
form
\begin{equation}
  h^2 Z_B Z_{B'} (W_1 - W_3\Phi_I),
\end{equation}
where plaquette $I$ is bounded by links $B$ and $B'$. We might say that $I$ 
is ``sandwiched'' in-between $B$ and $B'$.

The full second-order correction is
\begin{widetext}
\begin{eqnarray}
  H^{(2)}_\mu & = & 
  2h^2(S_1+T_1)\frac{N_\text{plaq}}{4}
  + h^2(S_2+T_2)\sum_{\ev{I,J}}\Phi_I\Phi_J
  - 4h^2(S_3+T_3)\sum_I\Phi_I \nonumber \\
  & & + h^2V_1\sum_{B\perp B'}Z_B Z_{B'}
  - h^2V_3\sum_{B\perp B'}Z_B Z_{B'} \Phi_{I_\text{elbow}}
  + h^2W_1\sum_{B\parallel B'}Z_B Z_{B'}
  - h^2W_3\sum_{B\parallel B'}Z_B Z_{B'} \Phi_{I_\text{sandwiched}},
  \qquad \label{H2solved}
\end{eqnarray}
\end{widetext}
where $\ev{I,J}$ denotes nearest-neighbor plaquettes $I$ and $J$. 
It is important to remark that all operators $Z_B Z_{B'}$ associated to hooks 
$B \perp B'$ are being summed over, even those that may be gauge-equivalent to 
others. The effective operators generated by renormalization are depicted in 
Fig.~\ref{fig:operators}.

Inspection of Eq.~(\ref{H2solved}) reveals five new gauge-invariant effective 
operators that have been generated by the renormalization transformation. The 
coefficients of these new operators will influence those obtained by successive
iterations of the decimation procedure. Therefore, we must go back and include 
these operators in $H_\sigma$. We choose to append them to $V_\sigma$ in 
Eq.~(\ref{Hseparated}) so that the eigenvalue problem on each cell remains
unchanged. Using the notation of Fig.~\ref{fig:operators},
\begin{equation}
  \text{add to $V_\sigma$} = \sum_{\alpha=1}^5 K_\alpha \mathcal{O}_\alpha
  + F N_\text{plaq}.
\end{equation}
As in Refs.~\onlinecite{HM} and \onlinecite{Hirsch} we treat the coefficients
$K_\alpha$ as being $O(h^2)$ since they are generated at second order in 
Hirsch--Mazenko perturbation theory. This means that we need only compute the 
analogue of Eq.~(\ref{H1}),
\begin{equation}
  \label{addtoHren}
  \text{add to $H_\mu^\text{ren}$} = 
  \sum_{i,i'}\ev{i'|(\text{add to $V_\sigma$})|i}\ket{\mu_i}\bra{\mu_{i'}}.
\end{equation}

For the identity operator, a single renormalization step reduces the number of 
plaquettes by a factor of 4. Therefore, $F' = 4F$.

Consider $\sum_{i,i'}\ev{i'|\mathcal{O}_\alpha|i}\ket{\mu_i}\bra{\mu_{i'}}$.
Generically, $\mathcal{O}_\alpha$ consists of a product of $Z$'s and $X$'s (or
$\sigma^z$'s and $\sigma^x$'s in the original Hilbert space). Depending on how 
this operator is situated on the lattice some of the spin operators will act on
internal links of a cell and some will act on external links. Since operators 
on different links commute, we can always write $\mathcal{O}_\alpha$ as a 
product of operators---one referring to internal links only and the other 
referring to external links only. Let $\mathcal{O}_\alpha = 
\mathcal{O}^\text{int}_\alpha \mathcal{O}^\text{ext}_\alpha$. Since internal 
links always belong to a specific cell, we may further decompose 
$\mathcal{O}^\text{int}_\alpha = \prod_I\mathcal{O}_\alpha^\text{int $I$}$,
where $I$ is the cell index. Then
\begin{eqnarray}
  \ev{i'|\mathcal{O}_\alpha|i} & = & 
  \prod_I \ev{0(\{x'\})|\mathcal{O}_\alpha^\text{int $I$}|0(\{x\})}_I 
  \nonumber \\
  & & \times
  \bra{x'}\dotsb\bra{x'}\mathcal{O}^\text{ext}_\alpha
  \ket{x}\dotsb\ket{x},
\end{eqnarray}
where all external links $x$ and $x'$ are involved in the matrix element. If we
now perform $\sum_{i'}$, which amounts to summing over all possible 
configurations of the external links $x'$, then any external link not directly
touched by $\mathcal{O}_\alpha^\text{ext}$ will receive a Kronecker delta
setting $x' = x$. Since $\ket{\mu_i}\bra{\mu_{i'}} = \prod \ket{x}\bra{x'}$,
$\ket{x}\bra{x'}$ corresponding to untouched links become $\ket{x}\bra{x}$.
Next, when $\sum_i$ is performed we might expect to get 
$\sum_{x = \pm}\ket{x}\bra{x} = \Id$ for those untouched external links. 
However, this is only true if another condition is met: those untouched 
external links must not lie on a cell that contains a touched link somewhere
else on its border or has $\mathcal{O}^\text{int}_\alpha$ acting on it. We shall
refer to such cells as ``touched'' cells. Otherwise, there will be a non-unit 
matrix element factor sitting inside the sum $\sum_{x = \pm}$. Our 
calculational algorithm can be stated as
\begin{eqnarray}
  & & \sum_{i,i'}\ev{i'|\mathcal{O}_\alpha|i}\ket{\mu_i}\bra{\mu_{i'}} 
  \nonumber \\ 
  & = &
  \underbrace{\sum_x\dotsc\sum_x}_{\substack{
  \text{directly touched}\\ \text{external links}}}
  \underbrace{\sum_x\dotsc\sum_x}_{\substack{
  \text{untouched}\\ \text{external links}\\ \text{on a touched cell}}} 
  \underbrace{\sum_{x'}\dotsc\sum_{x'}}_{\substack{
  \text{directly touched}\\ \text{external links}}}
  \prod_\text{touched cells $I$} \nonumber \\
  & &  
  \ev{0(\{x'\})|\mathcal{O}_\alpha^\text{int $I$}|0(\{x\})}_I\Bigr|_\text{$x' 
  = x$ for untouched links} \nonumber \\
  & & \times
  \underbrace{\bra{x'}\dotsb\bra{x'}}_{\substack{
  \text{directly touched}\\ \text{external links}}}
  \mathcal{O}^\text{ext}_\alpha
  \underbrace{\ket{x}\dotsb\ket{x}}_{\substack{
  \text{directly touched}\\ \text{external links}}} \nonumber \\
  & & \times 
  \underbrace{\ket{x}\bra{x'}\dotsb\ket{x}\bra{x'}}_{\substack{
  \text{directly touched}\\ \text{external links}}} \cdot
  \underbrace{\ket{x}\bra{x}\dotsb\ket{x}\bra{x}}_{\substack{
  \text{untouched}\\ \text{external links}\\ \text{on a touched cell}}}.
  \label{algorithm}
\end{eqnarray}

\begin{figure}
\includegraphics{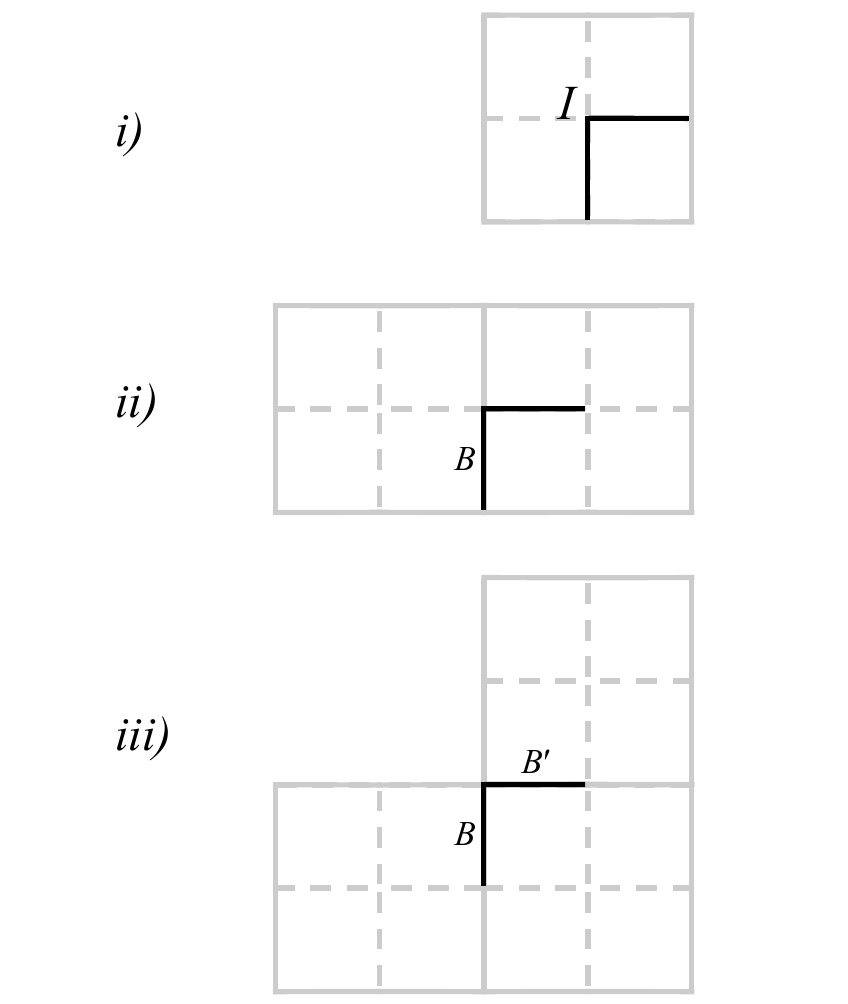}
\caption{\label{fig:hookcoupling}
How $\mathcal{O}_1$ generates effective operators.}
\end{figure}

Direct computation of expression (\ref{addtoHren}) is straightforward using 
Eq.~(\ref{algorithm}) but tedious. As an example, consider $\mathcal{O}_1$. 
There are three qualitatively distinct ways it can cover the original lattice.
See Fig.~\ref{fig:hookcoupling}. In Case (\textit{i}) the hook lies entirely 
on internal links. Define
\begin{equation}
  \chi = \ev{0(\{A\})|\sigma^z_{I,1,0}\sigma^z_{I,4,0}|0(\{A\}}_I.
\end{equation}
It turns out that $\chi$ is a function of the flux $\Phi_I = \prod_{k=1}^4 A_k$
only. Therefore, 
\begin{subequations}
\begin{eqnarray}
  \chi(+) & = & \ev{0_+|\sigma^z_{I,1,0}\sigma^z_{I,4,0}|0_+}_I, \\
  \chi(-) & = & \ev{0_-|\sigma^z_{I,1,0}\sigma^z_{I,4,0}|0_-}_I.
\end{eqnarray}
\end{subequations}
Thus,
\begin{eqnarray}
  \text{Case (\textit{i})} & = & 
  4K_1\frac{\chi(+)+\chi(-)}{2}\frac{N_\text{plaq}}{4} \nonumber \\
  & & + 4K_1\frac{\chi(+)-\chi(-)}{2}\sum_I \Phi_I.
\end{eqnarray}
There is a factor of 4 because there are four different orientations in which
the hook can lie entirely inside the plaquette. In Case (\textit{ii}) the hook
can lie in one of four ways on the shared boundary between two plaquettes. So 
\begin{equation}
  \text{Case (\textit{ii})} = 4K_1|\ev{0_-|0_+}_I|^2\sum_B Z_B.
\end{equation}
In Case (\textit{iii}) there is only a single orientation of the hook that 
straddles all three plaquettes. So
\begin{equation}
  \text{Case (\textit{iii})} = K_1|\ev{0_-|0_+}_I|^2\sum_{B\perp B'} Z_B Z_{B'}.
\end{equation}
The contributions from $\mathcal{O}_\alpha$ for $\alpha = 2,\dotsc,5$ may be 
worked out in a similar fashion.

Before writing the full expression for (\ref{addtoHren}) it will be convenient 
to adopt some shorthand notation. Internal-link spin operators, which are
normally denoted $\vec{\sigma}_{I,i,0}$, will be shortened to $\vec{\sigma}_i$. 
External-link spin operators will not appear explicitly in the matrix elements
so there is no chance of confusion. And we shall suppress the subscript $I$
on matrix elements. We find
\begin{widetext}
\begin{eqnarray}
  \text{add to $H_\mu^\text{ren}$} & = &
  \frac{N_\text{plaq}}{4} \Bigl[
  4F + 2K_1(\ev{0_+|\sigma^z_1\sigma^z_4|0_+} 
  + \ev{0_-|\sigma^z_1\sigma^z_4|0_-})
  + 2K_2(\ev{0_+|\sigma^z_1\sigma^z_2\sigma^x_1\sigma^x_2|0_+}
  + \ev{0_-|\sigma^z_1\sigma^z_2\sigma^x_1\sigma^x_2|0_-})
  \nonumber \\
  & & \qquad\qquad
  + 2K_5\bigl(\ev{0_+|\sigma^x_1\sigma^x_3|0_+} + \tfrac{1}{4}(
  \ev{0_+|\sigma^x_1\sigma^x_2|0_+}^2 + 2\ev{0_+|\sigma^x_1\sigma^x_2|0_+}
  \ev{0_-|\sigma^x_1\sigma^x_2|0_-} + \ev{0_-|\sigma^x_1\sigma^x_2|0_-}^2)\bigr)
  \Bigr] \nonumber \\
  & & + \sum_B Z_B \Bigl[
  4K_1|\ev{0_-|0_+}|^2 + 4K_2\ev{0_-|0_+}\ev{0_-|\sigma^x_1\sigma^x_2|0_+}
  + 4K_3|\ev{0_-|0_+}|^2 + 4K_4\ev{0_-|0_+}\ev{0_-|\sigma^x_1\sigma^x_2|0_+}
  \Bigr] \nonumber \\
  & & 
  + \sum_I\Phi_I\Bigl[
  2K_1(\ev{0_+|\sigma^z_1\sigma^z_4|0_+} - \ev{0_-|\sigma^z_1\sigma^z_4|0_-})
  + 2K_2(\ev{0_+|\sigma^z_1\sigma^z_2\sigma^x_1\sigma^x_2|0_+}
  - \ev{0_-|\sigma^z_1\sigma^z_2\sigma^x_1\sigma^x_2|0_-}) \nonumber \\
  & & \qquad\qquad
  +2K_5(\ev{0_+|\sigma^x_1\sigma^x_3|0_+}
  +\ev{0_+|\sigma^x_1\sigma^x_2|0_+}^2
  -\ev{0_-|\sigma^x_1\sigma^x_2|0_-}^2)\Bigr] \nonumber \\
  & & 
  + \sum_{B\perp B'}Z_BZ_{B'}\Bigl[
  K_1|\ev{0_-|0_+}|^2 + \half K_2 |\ev{0_-|0_+}|^2
  (\ev{0_+|\sigma^x_1\sigma^x_2|0_+} + \ev{0_-|\sigma^x_1\sigma^x_2|0_-})
  \Bigr] \nonumber \\
  & & 
  + \sum_{B\perp B'}Z_BZ_{B'}\Phi_{I_\text{elbow}}\Bigl[
  \half K_2 |\ev{0_-|0_+}|^2
  (\ev{0_+|\sigma^x_1\sigma^x_2|0_+} - \ev{0_-|\sigma^x_1\sigma^x_2|0_-})
  \Bigr] \nonumber \\
  & & 
  + \sum_{\ev{I,J}}\Phi_I\Phi_J\Bigl[
  \half K_5(\ev{0_+|\sigma^x_1\sigma^x_2|0_+}^2 
  - 2\ev{0_+|\sigma^x_1\sigma^x_2|0_+}\ev{0_-|\sigma^x_1\sigma^x_2|0_-}
  + \ev{0_-|\sigma^x_1\sigma^x_2|0_-}^2)
  \Bigr] \label{addtoHrensolved}
\end{eqnarray}
\end{widetext}
All matrix elements appearing above have been written in terms of $\ket{0_\pm}$
as given by Eqs.~(\ref{zeroplus}) and (\ref{zerominus}). It is worth noting 
that expression (\ref{addtoHrensolved}) with the alternative definition 
$\ket{0_-}_I = \ket{0(A_1 = -, A_2 = +, A_3 = +, A_4 = +)}_I$ would not be
correct.

The renormalized Hamiltonian to second order in the coupling $h$ is given by
adding expressions (\ref{H0solved}), (\ref{H1solved}), (\ref{H2solved}), and 
(\ref{addtoHrensolved}). This yields
\begin{widetext}
\begin{subequations}\label{recur}
\begin{eqnarray}
  H^\text{ren}_\mu & = & 
  - h'\sum_B Z_B - J'\sum_I\Phi_I \nonumber \\
  & & 
  + K_1'\sum_{B\perp B'}Z_B Z_{B'}
  + K_2'\sum_{B\perp B'}Z_B Z_{B'} \Phi_{I_\text{elbow}}
  + K_3'\sum_{B\parallel B'}Z_B Z_{B'}
  + K_4'\sum_{B\parallel B'}Z_B Z_{B'} \Phi_{I_\text{sandwiched}}
  + K_5'\sum_{\ev{I,J}}\Phi_I\Phi_J \nonumber \\
  & & 
  + F'\frac{N_\text{plaq}}{4}, \label{Hren}
\end{eqnarray}
where
\begin{eqnarray}
  h' & = & 2|\ev{0_-|0_+}|^2(h - 2K_1 - 2K_3) 
  - 4\ev{0_-|0_+}\ev{0_-|\sigma^x_1\sigma^x_2|0_+}(K_2 + K_4) \label{hrecur} \\
  J' & = & \frac{\epsilon^\text{c}_0(-)-\epsilon^\text{c}_0(+)}{2}
  + 4h^2(S_3+T_3) + 2K_1(\ev{0_-|\sigma^z_1\sigma^z_4|0_-} -
  \ev{0_+|\sigma^z_1\sigma^z_4|0_+}) \nonumber \\
  & & 
  + 2K_2(\ev{0_-|\sigma^z_1\sigma^z_2\sigma^x_1\sigma^x_2|0_-} - 
  \ev{0_+|\sigma^z_1\sigma^z_2\sigma^x_1\sigma^x_2|0_+}) \nonumber \\
  & & 
  + 2K_5(\ev{0_-|\sigma^x_1\sigma^x_2|0_-}^2
  - \ev{0_+|\sigma^x_1\sigma^x_3|0_+}
  - \ev{0_+|\sigma^x_1\sigma^x_2|0_+}^2) \label{Jrecur} \\
  K_1' & = & h^2V_1 + |\ev{0_-|0_+}|^2\bigl(K_1 + \half K_2(
  \ev{0_+|\sigma^x_1\sigma^x_2|0_+} + \ev{0_-|\sigma^x_1\sigma^x_2|0_-})
  \bigr) \label{K1recur} \\
  K_2' & = & -h^2V_3 + \half K_2 |\ev{0_-|0_+}|^2
  (\ev{0_+|\sigma^x_1\sigma^x_2|0_+} - \ev{0_-|\sigma^x_1\sigma^x_2|0_-}) 
  \label{K2recur} \\
  K_3' & = & h^2W_1 \label{K3recur} \\
  K_4' & = & -h^2W_3 \label{K4recur} \\
  K_5' & = & h^2(S_2+T_2) + 
  \half K_5(\ev{0_+|\sigma^x_1\sigma^x_2|0_+}-
  \ev{0_-|\sigma^x_1\sigma^x_2|0_-})^2 \label{K5recur} \\
  F' & = & \frac{\epsilon_0^\text{c}(+)+\epsilon_0^\text{c}(-)}{2} + 
  2h^2(S_1+T_1) + 4F
  + 2K_1(\ev{0_+|\sigma^z_1\sigma^z_4|0_+} 
  + \ev{0_-|\sigma^z_1\sigma^z_4|0_-}) \nonumber \\
  & &
  + 2K_2(\ev{0_+|\sigma^z_1\sigma^z_2\sigma^x_1\sigma^x_2|0_+}
  + \ev{0_-|\sigma^z_1\sigma^z_2\sigma^x_1\sigma^x_2|0_-})
  \nonumber \\
  & &
  + 2K_5\bigl(\ev{0_+|\sigma^x_1\sigma^x_3|0_+} + \tfrac{1}{4}(
  \ev{0_+|\sigma^x_1\sigma^x_2|0_+} + \ev{0_-|\sigma^x_1\sigma^x_2|0_-})^2
  \bigr). \label{Frecur}
\end{eqnarray}
\end{subequations}
\end{widetext}
The matrix elements appearing above may be calculated using the cell-basis 
representations for internal $\sigma^z$ matrices given in 
Eqs.~(\ref{internalZ}), and 
\begin{subequations}
\begin{eqnarray}
  \sigma^x_1\sigma^x_2 & = & 
  \begin{pmatrix}
  & & & & 0 & 1 & 0 & 0 \\
  & & & & 0 & 0 & 0 & 1 \\
  & & & & 0 & 0 & 1 & 0 \\
  & & & & 1 & 0 & 0 & 0 \\
  0 & 0 & 0 & 1 & & & & \\
  1 & 0 & 0 & 0 & & & & \\
  0 & 0 & 1 & 0 & & & & \\
  0 & 1 & 0 & 0 & & & &
  \end{pmatrix}, \\
  \sigma^x_1\sigma^x_3 & = & 
  \begin{pmatrix}
  0 & 0 & 1 & 0 & & & & \\
  0 & 0 & 0 & 1 & & & & \\
  1 & 0 & 0 & 0 & & & & \\
  0 & 1 & 0 & 0 & & & & \\
  & & & & 0 & 0 & 0 & 1 \\
  & & & & 0 & 0 & 1 & 0 \\
  & & & & 0 & 1 & 0 & 0 \\
  & & & & 1 & 0 & 0 & 0
  \end{pmatrix}.
\end{eqnarray}
\end{subequations}

Eqs.~(\ref{recur}) are recursion relations for the operator coefficients in the
Hamiltonian. We have checked that they are equivalent to the recursion relations
derived by Hirsch in Ref.~\onlinecite{Hirsch} for the quantum Hamiltonian of 
the square-lattice Ising model in a transverse magnetic field. \footnote{Our 
couplings $h$, $J$, $K_1$, $K_2$, $K_3$, $K_4$, $K_5$, and $F$ are dual to 
Hirsch's couplings $\Delta$, $\epsilon$, $-\mu/2$, $\alpha$, $-\delta$, 
$\lambda$, $-\beta$, and $d$, respectively. See Ref.~\onlinecite{Hirsch}.
Minus signs account for different conventions used
to define renormalized operators. The reason for a factor of a half in 
$K_1 = -\mu/2$ is because $Z_{B_1}Z_{B_2}$ (where $B_1 \perp B_2$) 
is gauge-equivalent to $Z_{B_3}Z_{B_4}$ if $B_1$, $B_2$, $B_3$, and $B_4$ 
are four links meeting at the same site. It should be also noted that errant 
signs and minor typos exist in Hirsch's recursion relations. See Eqs.~(A1) and
(A2) in the appendix of Ref.~\onlinecite{Hirsch}. A recomputation of the 
renormalized Hamiltonian in the tranverse field Ising model reveals the 
following corrections. In the equation for 
$\epsilon_{n+1}$, all instances of $\beta_n$ should have the opposite sign as 
the one written. In the equation for $\beta_{n+1}$, there should be a minus 
sign instead of a plus sign in front of the $\tfrac{1}{4}$. In the expression
for $B_2$, $1/(E_0-E_n)$ ought to be $1/(E_1-E_n)$. In the expression for 
$C_2$, $(n,n') \neq (0,1)$ ought to read $(n,n') \neq (1,0)$. Correcting these
typos does not change any of the conclusions of that paper.} This check is 
accomplished using a duality transformation. \cite{Kogut} The
't Hooft disorder operator, given by a string of $\sigma^z$'s in the lattice 
gauge theory, corresponds to the order parameter, $\sigma^x$, in the Ising 
model. And the magnetic flux operator, given by a product of $\sigma^x$'s around
a plaquette, maps to the transverse field operator, $\sigma^z$, living at the
site dual to the plaquette.
 
It is certainly more challenging to obtain the recursion relations in the
lattice gauge theory than in the Ising model---gauge invariance enlarges the 
number of possible states, and additional formalism is necessary to restrict 
to the gauge-invariant sector. For instance, in the former one is forced to 
consider the cellular magnetic flux $\Phi_I$ as potentially dependent on eight 
boundary conditions, whereas in the Ising model no such boundary conditions are
required. The fact that our results agree with those obtained from
the simpler (non-gauged) formulation of the theory is a reassuring check that 
we correctly implemented the Hirsch--Mazenko procedure.

The renormalized Hamiltonian given by expression (\ref{Hren}) maintains gauge
invariance on the thinned square lattice which has a quarter as many plaquettes
as the original lattice. Local gauge transformations are defined by operators 
associated to the sites or vertices between links. At any site $\vec{r}$ 
in the thinned lattice the generator is
\begin{equation}
  G_{\vec{r}} = \prod_{\substack{\text{links $B$ emerging}\\ 
  \text{from $\vec{r}$}}} Z_B.
\end{equation}
$G_{\vec{r}}$ commutes with $H^\text{ren}_\mu$.

\subsection{Critical point, fixed points, and eigenvalues}

Let us analyze the recursion relations given by Eqs.~(\ref{recur}). Treating
$h$ as an energy scale, we shall work with dimensionless couplings grouped
into the tuple $(J/h, K_1/h, \dotsc, K_5/h)$. This quantity, which we may 
abbreviate as $(J/h,\vec{K}/h)$ is a six-dimensional real-valued vector. By
starting at an arbitrary vector, iterations of the recursion relations in the 
form
\begin{equation}\label{flow}
  (J/h, \vec{K}/h) \mapsto (J'/h', \vec{K'}/h')
\end{equation}
yield a sequence of points which describe a ``flow'' of the Hamiltonian defined
over successively thinner lattices. This flow should preserve the low-energy 
spectrum of the original lattice Hamiltonian. Note that Eq.~(\ref{flow}) is
obtained by dividing Eqs.~(\ref{Jrecur}) through (\ref{K5recur}) by 
Eq.~(\ref{hrecur}). 

\begin{figure}
\includegraphics{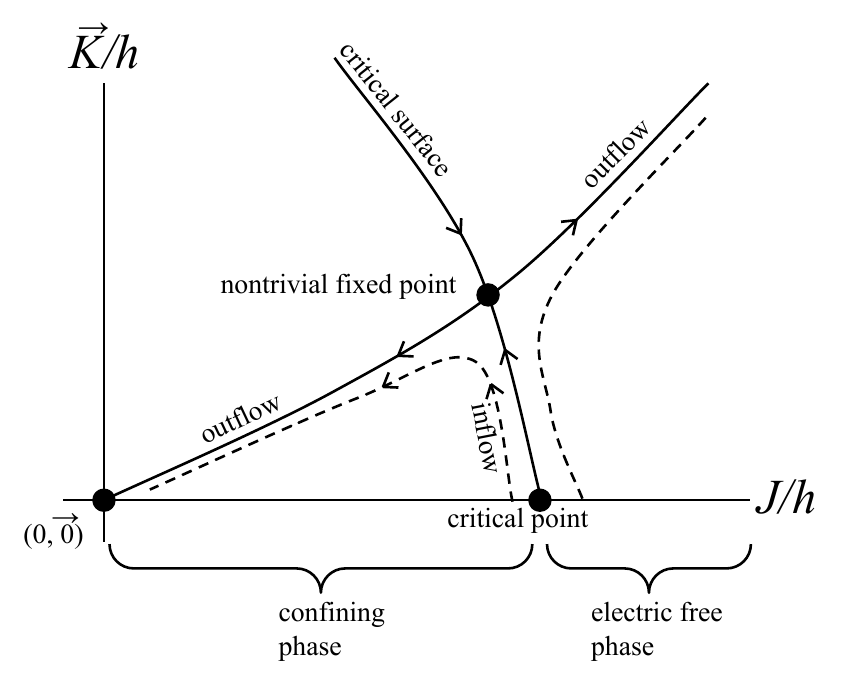}
\caption{\label{fig:flow}
A heuristic picture of the sequence of iterations of the recursion relations
visualized as a flow in the six-dimensional space of dimensionless couplings.
$\vec{K}/h$ represents a vector of couplings $K_\alpha/h$ for 
$\alpha = 1,\dotsc,5$. The critical surface has codimension 1. The outflow 
trajectory from the nontrivial fixed point to the origin is a line.}
\end{figure}

We study numerically flows that begin on the Ising axis $(J/h, \vec{0})$. Let
\begin{equation}
  (J/h)_\text{c} = 3.56895.
\end{equation}
For $J/h > (J/h)_\text{c}$, flows have the property
that $|J/h|$ grows without bound. And for $J/h < (J/h)_\text{c}$, flows 
approach the origin $(0, \vec{0})$. Therefore, $((J/h)_\text{c},\vec{0})$ 
is the critical point, i.e., the intersection of the critical surface with 
the Ising axis. A flow near this point will initially approach (along an 
``inflow'') a nontrivial fixed point before veering away (hugging the 
``outflow'') toward the stable fixed points at infinity and the origin. See 
Fig.~\ref{fig:flow}.

The nontrivial and unstable fixed point is found by applying Newton's method to
the beta function $(J'/h',\vec{K'}/h') - (J/h,\vec{K}/h)$. It is located at
\begin{eqnarray}
\label{fixedpt}
  (J/h,\vec{K}/h)_* & = & 
  (2.72662,\,-0.44280,\,0.41555,\nonumber \\
  & & \quad -0.16793,\,0.46845,\,-0.22900). \qquad\quad
\end{eqnarray}
Our results corroborate those obtained by Hirsch for the quantum Ising model
in a transverse field. \footnote{Hirsch's fixed point is given by Eq.~(9) in 
Ref.~\onlinecite{Hirsch}, but we believe there is a typo: the values for 
$\lambda/\Delta$ and $\delta/\Delta$ should be swapped. His critical coupling 
is given by Eq.~(10) and it is $3.556$, which is slightly smaller than our 
value.}

Renormalization for flows in the vicinity of the nontrivial fixed point are
particularly simple since the recursion relations may be linearized. Let us
denote the vector $(J/h,\vec{K}/h)$ more succintly by $\kappa$. Then
\begin{equation}
  \kappa' \simeq \kappa_* + R(\kappa - \kappa_*),
\end{equation}
where $R = \partial\kappa'/\partial\kappa|_{\kappa = \kappa_*}$ is the 
Jacobian matrix of partial derivatives evaluated at the fixed point. Our matrix
$R$ turns out not to be symmetric and therefore, not all eigenvalues are 
guaranteed to be real. The eigenvalues are (ordered from largest to smallest 
magnitude):
\begin{eqnarray}
  \Lambda & = & \{2.89173,\,0.41344,\,0.24637+0.15397\mathrm{i}, \nonumber \\
  & & \quad 0.24637-0.15397\mathrm{i},\,0.08319,\,0\}.
\end{eqnarray}
Note that there is a complex conjugate pair of eigenvalues. This strange 
feature was noted by Hirsch and it seems to indicate an inconsistency in the 
renormalization-group equations \cite{Hirsch}. Notwithstanding this blemish, if
the left eigenvector corresponding to $\Lambda_k$ is denoted $e_k$, then we may
form scaling variables by taking a dot product: $u_k = e_k\cdot\kappa$. These
scaling variables renormalize multiplicatively, i.e., $u_k' = \Lambda_k u_k$.
Since $\Lambda_2,\dotsc, \Lambda_6$ all have absolute value less than 1, the
scaling variables $u_2,\dotsc,u_6$ renormalize to zero. These five coordinates
correspond to the irrelevant directions along the critical surface that guide
flows into the nontrivial fixed point. However, since $\Lambda_1 > 1$, the
scaling variable $u_1$ is relevant and iterations of the recursion relations
will tend to make this coordinate grow. Thus, the eigenvector $e_1$ must define
the outflow trajectory in the linear space around the nontrivial fixed point.

\bibliography{quantrg.bib}

\end{document}